\documentclass[aps,prl,twocolumn,superscriptaddress,longbibliography]{revtex4-1}
\usepackage{amsmath,amsthm,amssymb,amsfonts,bbold}
\usepackage{graphicx,float}
\usepackage[font=small]{caption}
\usepackage{subcaption}
\graphicspath{{Free_boson/pics/}}
\usepackage{tikz}
\usepackage{multirow}
\usepackage[colorlinks]{hyperref}

\newcommand{\figref}[1]{Fig.~\ref{#1}}

\newcommand{\tabref}[1]{Table~\ref{#1}}
\newcommand{\ket}[1]{|#1\rangle}

\newenvironment{eqns}
{\begin{equation}\begin{aligned}}
{\end{aligned}\end{equation}} 

\begin{document}

\title{Fixed-point tensor network for compactified boson conformal field theory}
\author{Gong Cheng}
\affiliation{Department of Physics, Virginia Tech, Blacksburg, VA 24060, USA}
\affiliation{Virginia Tech Center for Quantum Information Science and Engineering, Blacksburg, VA 24061,
USA}
\author{Dong-Yu Bao}
\affiliation{Department of Physics, The Chinese University of Hong Kong, Sha Tin, New Territories, Hong Kong, China}
\author{Zheng-Cheng Gu}
\email{zcgu@phy.cuhk.edu.hk}
\affiliation{Department of Physics, The Chinese University of Hong Kong, Sha Tin, New Territories, Hong Kong, China}

\begin{abstract}
Fixed-point (FP) tensor networks provide a discrete spacetime representation of conformal field theories (CFTs), offering a new route toward understanding holographic duality, generalized symmetries, and even quantum gravity. In this work, we construct FP tensors for the 2D compactified boson theory at a generic compactification radius—an archetypal irrational CFT—using boundary (open‑string) data with conformal boundary conditions. We show that the resulting tensors reproduce the closed‑string spectrum with high accuracy and generate stable renormalization‑group (RG) flows under the tensor complex renormalization algorithm. Moreover, we identify a controllable exactly marginal deformation at the level of a single tensor, enabling flows that move continuously along the $c=1$ moduli space. This framework establishes a concrete lattice‑level route toward describing a broad class of 2D irrational CFTs.
\end{abstract}
\maketitle


\emph{Introduction} — The two-dimensional compactified boson theory is one of the most fundamental conformal field theories (CFTs). It plays a central role across several areas of theoretical physics, including string theory \cite{Klebanov:1991qa,Polchinski:1998rq}, condensed matter physics \cite{Minic:1992yb,Degiovanni:1997re}, and statistical mechanics \cite{PhysRevResearch.4.023159,PhysRevE.101.060105}. It not only provides the universal low-energy description of a wide class of gapless one-dimensional quantum systems, such as Luttinger liquids \cite{luttingerExactlySolubleModel1963, Haldane_1981}, spin chains \cite{Affleck_1989}, and superfluids \cite{Cazalilla_2011}), but also serves as a basic building block of worldsheet string theory, where compactification gives rise to interesting physics such as T-duality and enhanced symmetry at special radii \cite{ginsparg1988}. Despite its simple free-field formulation, the compactified boson possesses a remarkably rich structure: it constitutes a continuous family of CFTs parametrized by the compactification radius, which is generically irrational, while admitting special rational points with enlarged chiral symmetry. For these reasons, it serves as a canonical laboratory for exploring broad questions in CFT.

Recent developments in the topological holographic principle and generalized-symmetry \cite{PhysRevResearch.2.033417, PhysRevResearch.2.043086, Kong:2020jne, Freed:2022qnc, PhysRevB.107.155136,Kong2019AMT} description of CFT suggest that continuous quantum field theories admit an intrinsically algebraic reformulation in discrete spacetime. Motivated by this perspective, a recent advance introduced an exact fixed-point (FP) tensor construction
for rational CFTs (RCFTs) \cite{chengPrecisionReconstructionRational2025}.  In this framework, spacetime is discretized using a basis of boundary changing operators (BCOs), organized into representations of the Virasoro algebra. The basic building blocks of the FP tensor are correlation functions of these BCOs, which naturally glue together into a spacetime tensor-network representation of the full CFT path integral.  Moreover, the FP tensor admits a natural decomposition into topological data encoding generalized symmetry and geometric data encoding conformal symmetry. This structure makes the algebraic reorganization of the quantum field theory manifest.

Although the FP tensor construction was shown to reconstruct RCFT path integrals with high precision, especially when combined with tensor network renormalization (TNR) \cite{PhysRevLett.99.120601,bao2025tensorcomplexrenormalizationgeneralized,guTensorentanglementfilteringRenormalizationApproach2009,yangLoopOptimizationTensor2017}, it remains unclear whether this framework extends beyond the rational setting. The central obstacle is the irrational nature of the theory: fusion typically generates infinitely many representations, while the available symmetry is less constraining relative to the complexity of the spectrum. As a result, the algebraic structure becomes substantially less tractable. In boundary CFTs (BCFTs) of irrational theories, for example, the set of representations contributing to Ishibashi states is typically infinite or even continuous, so boundary states are expressed as infinite sums or integrals rather than finite discrete combinations. This makes a fully discrete formulation much more challenging than in the rational case. 

In this Letter, we overcome these difficulties and extend the FP tensor construction beyond RCFTs by studying the compactified free boson at a generic compactification radius. 
We show that this irrational CFT nevertheless admits a fully discrete spacetime-lattice description as well. We further identify a controllable exactly marginal deformation at the level of a single tensor.
We stress that although tensor network methods have previously been applied to bosonic field theories~\cite{Shimizu:2012zza, Shimizu:2012wfa, Campos:2019, Campos:2021, Hu:2018hyd} and other lattice field theories~\cite{Kadoh:2018, Meurice:2019ddf, Bazavov:2019qih, Butt_2020, Kadoh:2022}, here we adopt a different approach, constructing exact FP tensor networks for CFTs from generalized symmetry.


\emph{Fixed-point tensor construction} — The Euclidean path integral of an RCFT on a surface can be represented as a spacetime tensor network 
\begin{align}
Z_{M} =  \sum_{\{ (i,I)\}, \{a\} } \prod_v \omega_{a}    
\prod_{\triangle}\mathcal{T}^{a b c}_{(i,I) (j,J)(k,K)}.  \label{eq:partition}
\end{align}
Here, $\triangle$ denotes an elementary triangle in a triangulation of the surface, and $v$ denotes vertices. The local building
block $\mathcal{T}^{abc}_{(i,I)(j,J)(k,K)}$ is called a FP tensor \cite{chengPrecisionReconstructionRational2025}. It is invariant
under TNR and therefore represents the renormalization group (RG) fixed point of the field theory.

The FP tensor is constructed from boundary three-point functions of BCOs in the associated BCFT. Each BCO is labeled by
a pair $(i,I)$, where $i$ denotes the conformal family and $I$ labels a descendant within that family.
A BCO changes the conformal boundary condition (CBC) from $a$  to $b$, and therefore exists only
when the corresponding representation appears in the fusion product, $i\in a\times b$. Each CBC $a$
is assigned a weight $\omega_a$, chosen such that the resulting superposition of boundary conditions is
transparent to topological defects \cite{Brehm:2021wev,PhysRevD.104.026012}.

\begin{figure}[tb]
    \centering
    \includegraphics[width=0.9\linewidth]{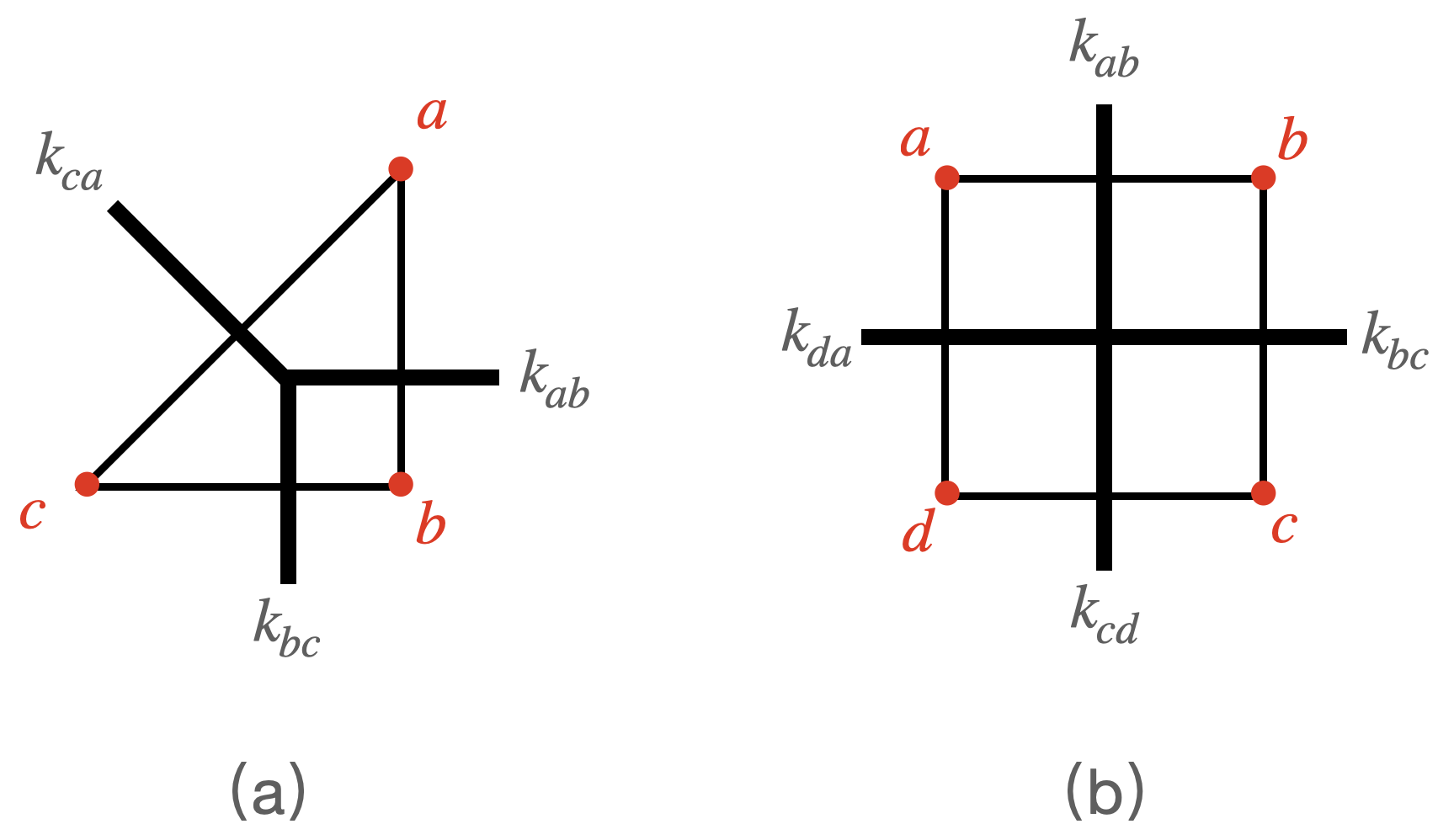}
    \caption{Fixed-point tensor: (a) rank-3 tensor. (b) rank-4 tensor.}
    \label{fig:rank3}
\end{figure}

We now extend this construction to the compactified boson CFT with central charge $c=1$. The
basic geometry of the rank-3 tensor is shown in Fig.~\ref{fig:rank3} (a). The degrees of freedom on each
tensor leg are open-string vertex operators. We label the $U(1)$ charge carried by an operator changing the boundary condition from $a$ to $b$ by $k_{ab}$. The CBCs may be chosen
to be Dirichlet boundary conditions, labeled by the D-brane position
$x_a\in[0,2\pi R)$, or Neumann boundary conditions, labeled by the Wilson line
$\theta_a\in[0,\pi/R)$. 

For a three-point function involving boundary conditions $a,b,c$, we use the shorthand
\begin{eqns}
    k_1\equiv k_{ab}, \quad k_2\equiv k_{bc}, \quad k_3\equiv k_{ca}. 
\end{eqns}%
Charge conservation then requires 
$k_1+k_2+k_3=0$, and the boundary three-point function is
\begin{eqns}\label{eq:VVV}
    &\langle V_{k_1}^{ab}(z_1)V_{k_2}^{bc}(z_2)V_{k_3}^{ca}(z_3)\rangle\\
    =&|z_1-z_2|^{k_1\cdot k_2}|z_2-z_3|^{k_2\cdot k_3}|z_3-z_1|^{k_3\cdot k_1}.
\end{eqns}%
The allowed values of the charge $k_{ab}$ depend on the boundary conditions. For Dirichlet boundary conditions,
\begin{eqns}\label{eq:D-Dk}
    k_{ab}^{(D-D)}=2\left(\frac{x_a-x_b}{2\pi}+w_{ab}R\right), 
\end{eqns}%
where $w_{ab}\in \mathbb{Z}$ is the winding number of the open-string stretching between the D-branes at $x_a$ and $x_b$.  For Neumann boundary conditions, the allowed charges are 
\begin{eqns}\label{eq:N-Nk}
    k_{ab}^{(N-N)}&=\frac{\theta_a-\theta_b}{\pi}+\frac{n_{ab}}{R}.
\end{eqns}%
where $n_{ab}\in \mathbb{Z}$ is the quantized momentum number of the open string in the compact direction. The corresponding conformal weights are
\begin{eqns}\label{eq:kab_DD}
   h_{ab}=&\frac{1}{2}k_{ab}^2. 
\end{eqns}%

To obtain the FP tensor, we map the boundary three-point function to the isosceles right-triangle geometry
shown in Fig.~\ref{fig:rank3} (a). Let $\chi_i$ denote the local coordinate map associated with the $i$th
leg of the triangle. For compactness, we write the primary field $V_{k_i}^{ab}$ as $V_i$, with conformal weight $h_i\equiv h_{ab}$. Conformal covariance gives 
\begin{eqns}
\chi_{i*}V_i(z_i)=\bigl|\chi_i'(z_i)\bigr|^{h_i}\,V_i\bigl(\chi_i(z_i)\bigr).
\end{eqns}%
Therefore, the three-point function transforms as
\begin{eqns}
    \langle &\chi_{1*}V_1(z_1)\chi_{2*}V_2(z_2)\chi_{3*}V_3(z_3)\rangle\\
    =&\frac{|\chi_1'|^{h_1}|\chi_2'|^{h_2}|\chi_3'|^{h_3}}{|z_1-z_2|^{-k_1\cdot k_2}|z_2-z_3|^{-k_2\cdot k_3}|z_3-z_1|^{-k_3\cdot k_1}}.
\end{eqns}%
The explicit forms of the maps $\chi_i$ for the triangular geometry were derived in 
Ref.~\cite{chengPrecisionReconstructionRational2025} and are summarized in the Supplemental Material. 

When only primary fields are retained on each tensor leg, the FP tensor elements take the simple form
\begin{eqns}\label{eq:primarytensor}
    T_{h_1h_2h_3}^{abc}=C_{h_1h_2h_3}^{abc}p^{h_1+h_2}q^{h_3},
\end{eqns}%
where the geometric factors for the isosceles right triangle are
$p\approx0.26658$ and $q\approx0.70421$. 
The coefficients $C_{h_1h_2h_3}^{abc}$ are the boundary three-point structure constants appearing in Eq.~\eqref{eq:VVV}. For the free-boson vertex operators considered here, these structure constants are unity, so the tensor elements are determined entirely by the conformal weights and the geometric factors.

Having discussed the BCOs and tensor elements, we now focus on the CBCs $a,b,c$. In the compactified boson CFT, the Dirichlet and Neumann CBCs are labeled by
continuous parameters: the D-brane position $x$ and the Wilson line $\theta$, respectively.The corresponding weights $\omega_{a,b,c}$ are chosen to satisfy the entanglement brane condition~\cite{PhysRevD.104.026012}. To construct a valid FP tensor, it is sufficient to work entirely within either the Dirichlet or 
Neumann family of boundary conditions. The two resulting tensor networks are related by T-duality, $R\rightarrow 1/(2R)$. In either case, the tensor carries a continuous boundary-condition label, so the contraction in Eq.~\eqref{eq:partition} becomes an integral over $x$ or $\theta$, giving rise to a continuous tensor network formulation, see End Matter for more details.  To obtain a finite-dimensional tensor suitable for numerical calculations, we replace these continuous families by finite grids:
\begin{eqns}
    x_m&=\frac{m}{\Lambda_D} 2\pi R, \qquad m=0, 1,2, \cdots \Lambda_D-1\\
    \theta_l&=\frac{l}{\Lambda_N}\frac{\pi}{R}, \qquad l=0,1,2, \cdots \Lambda_N-1,
\end{eqns}%
with the grid sizes $\Lambda_D$ and $\Lambda_N$ satisfying the
cutoff conditions
\begin{eqns}\label{eq:Lambdas}
    \Lambda_D>2\sqrt{2}R,\qquad\Lambda_N>\frac{\sqrt{2}}{R}.
\end{eqns}%
These cutoff conditions, derived in the End Matter, ensure that the superposition of the Cardy states over either grid selects the vacuum Ishibashi state together with a tower of irrelevant higher-charge primaries:
\begin{eqns}\label{eq:DNcardy}
    \frac{1}{\Lambda_D}\sum_{m=0}^{\Lambda_D-1} \ket{D,x_m}_c&=(2R)^{-1/2}\left(\ket{0}+J_{-1}\bar J_{-1}\ket{0}+\cdots\right),\\
    \frac{1}{\Lambda_N}\sum_{l=0}^{\Lambda_N-1}\ket{N, \theta_l}_c&=R^{1/2}\left(\ket{0}-J_{-1}\bar J_{-1}\ket{0}+\cdots\right),
\end{eqns}%
where the ellipsis denotes irrelevant contributions. Note that with a single set of CBCs, the finite bond-dimension cutoff of the FP tensor generically induces an effective marginal deformation due to the presence of the level-(1,1) contribution $J_{-1}\bar J_{-1}\ket{0}$ in Eq.~\eqref{eq:DNcardy}.  However, this marginal deformation is suppressed as more descendant states are retained in the FP tensor, as illustrated below. 

\begin{figure}[tb]
    \centering
    \includegraphics[width=0.9\linewidth]{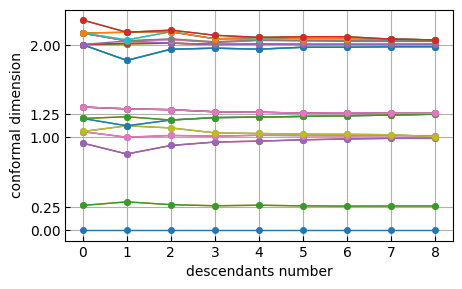}
    \caption{Closed-string conformal dimension from rank-3 FP tensor with growing descendants cutoff at $R=1$.}
    \label{fig:desendants}
\end{figure}
 
\emph{Numerical verification} — 
Having constructed the FP tensors from open-string data, we now test the validity of the construction by extracting the closed-string spectrum from the associated transfer matrix, following the method of Ref.~\cite{chengPrecisionReconstructionRational2025}. This provides a more stringent test than reproducing the partition function. The transfer-matrix construction proceeds in two steps. First, rank-3 tensors are contracted to form rank-4 tensors. These rank-4 tensors are then contracted into a length-three cylinder. Details of the transfer-matrix construction are given in the Supplemental Material.
The resulting eigenvalues are compared with the exact closed-string spectrum of the compact boson
\begin{eqns}
    \Delta=&(\frac{n}{2R})^2+m^2R^2+A+\bar A,
\end{eqns}%
where $A$ and $\bar A$ denote the descendant levels in the holomorphic and antiholomorphic $U(1)$ Kac-Moody sectors, respectively.

In the numerical evaluation, the $U(1)$ charges $w$ and $n$ in Eq.~\eqref{eq:D-Dk} and Eq.~\eqref{eq:N-Nk} are truncated to finite sets $\{-w_{\max},...,w_{\max}\}$ and $\{-n_{\max},...,n_{\max}\}$, respectively. Nevertheless, the closed-string spectrum is recovered with high accuracy. \figref{fig:desendants} show the spectra extracted from the rank-3 FP tensors at the radius
$R=1$ (the free Dirac fermion point). We observe systematic convergence of the numerical spectrum to the exact closed-string spectrum as the number of descendant states retained in the FP tensor increases.

\begin{figure}[tb]
    \centering
    \includegraphics[width=.85\linewidth]{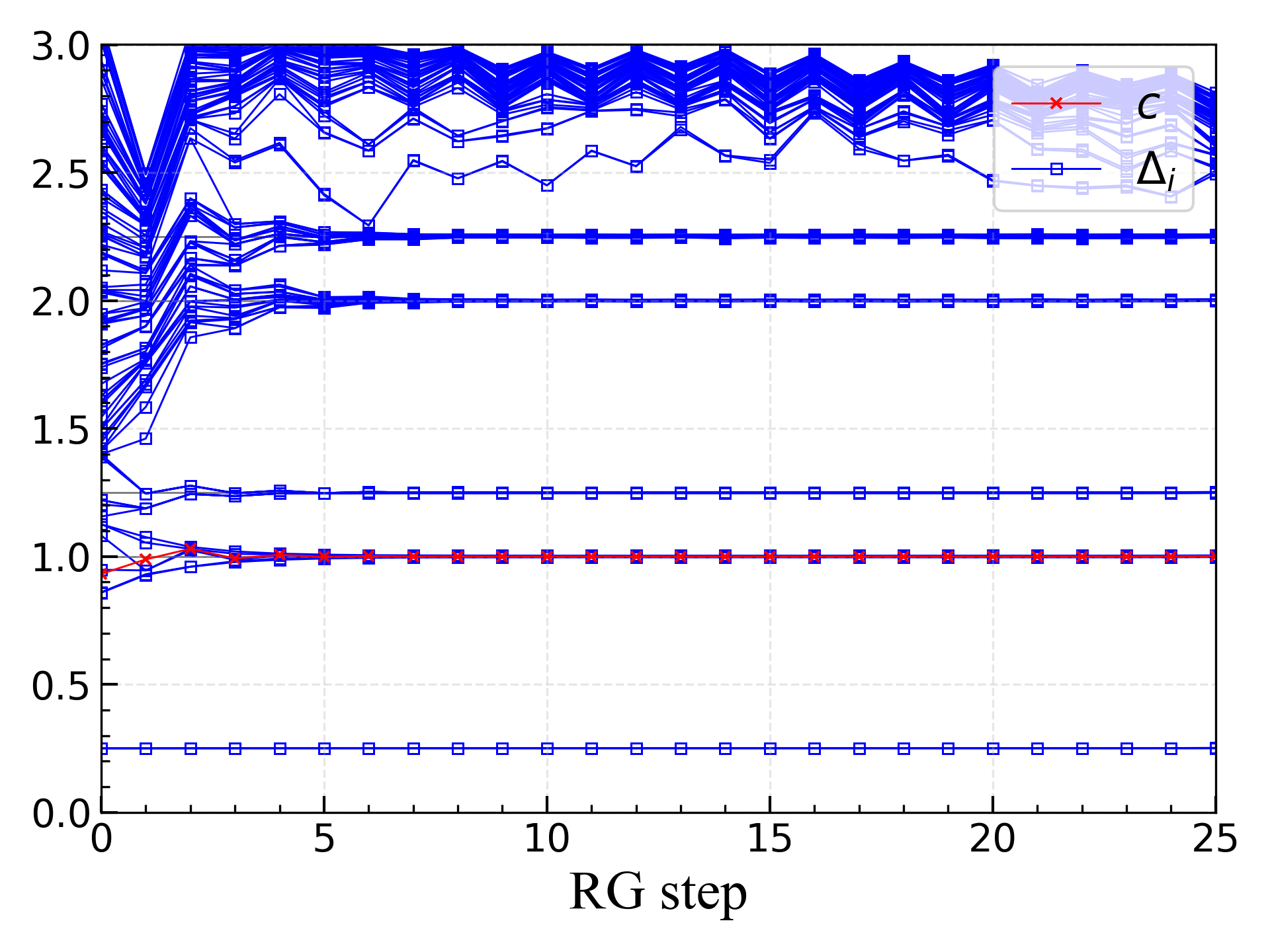}
    \caption{RG spectrum obtained from the refined rank-3 FP tensor at $R=1$, $\Lambda_D=4$, $\Lambda_N=2$, $w_{\max}=n_{\max}=2$, $\chi=20$.}
    \label{fig:3pt_DN_disk_1.000}
\end{figure}

\emph{Refined FP tensor}  —
At finite truncation, a FP tensor built from a single boundary-condition family generically retains a residual marginal perturbation Eq. ~\eqref{eq:DNcardy}, which will change the radius $R$ by performing TNR based algorithm. Here we construct a \textit{refined FP tensor} in which this perturbation is canceled. To eliminate such a marginal perturbation, we introduce a \emph{mixed shrinkable boundary condition} (MSBC), in which both types of boundary conditions are included in the sum in Eq.~\eqref{eq:partition}. The Dirichlet and Neumann boundary conditions are assigned weights 
\begin{eqns}
    \omega_D=(2R)^{1/2}/\Lambda_D, \qquad \omega_N=1/(R^{1/2}\Lambda_N),  
\end{eqns}%
respectively. The Cardy state associated with the MSBC is then the weighted linear combination of the discretized Dirichlet and Neumann Cardy states:
\begin{eqns}
&\ket{\mathcal{B}_{\rm mix}}\\
:=
    &\frac{1}{2}\left(\frac{(2R)^{1/2}}{\Lambda_D}\sum_{m=0}^{\Lambda_D-1} \ket{D,x_m}_c+\frac{1}{R^{1/2}\Lambda_N}\sum_{l=0}^{\Lambda_N-1}\ket{N,\theta_l}_c\right).
\end{eqns}%
In this symmetric combination, the level-$(1,1)$ contributions from the Dirichlet and Neumann sectors enter with opposite signs and therefore cancel each other. 

We now adopt the MSBC in the FP tensor construction, thereby defining the refined FP tensor. In this construction, we introduce additional BCOs, namely twist fields, that change Dirichlet boundary conditions into Neumann boundary conditions, and vice versa. Their three-point functions and the corresponding tensor elements are given in the End Matter. 

We numerically verify the refined rank-3 FP tensor with MSBC using TCR algorithm \cite{bao2025tensorcomplexrenormalizationgeneralized}. The input tensor contains only primary states and is truncated to its lowest-lying components with a relatively small cutoff $w_{\max}$ and $n_{\max}$. The extracted central charge and conformal dimensions at $R=1$ are shown in Fig.~\ref{fig:3pt_DN_disk_1.000}, where $\chi$ denotes the TCR bond dimension. The RG flow remains stable and accurately reproduces the expected conformal spectrum, with improved agreement with the exact results and the correct degeneracy structure. Detailed TCR procedures and results at additional radii are provided in the Supplemental Material.






\begin{figure}[tb]
    \centering
    \includegraphics[width=.85\linewidth]{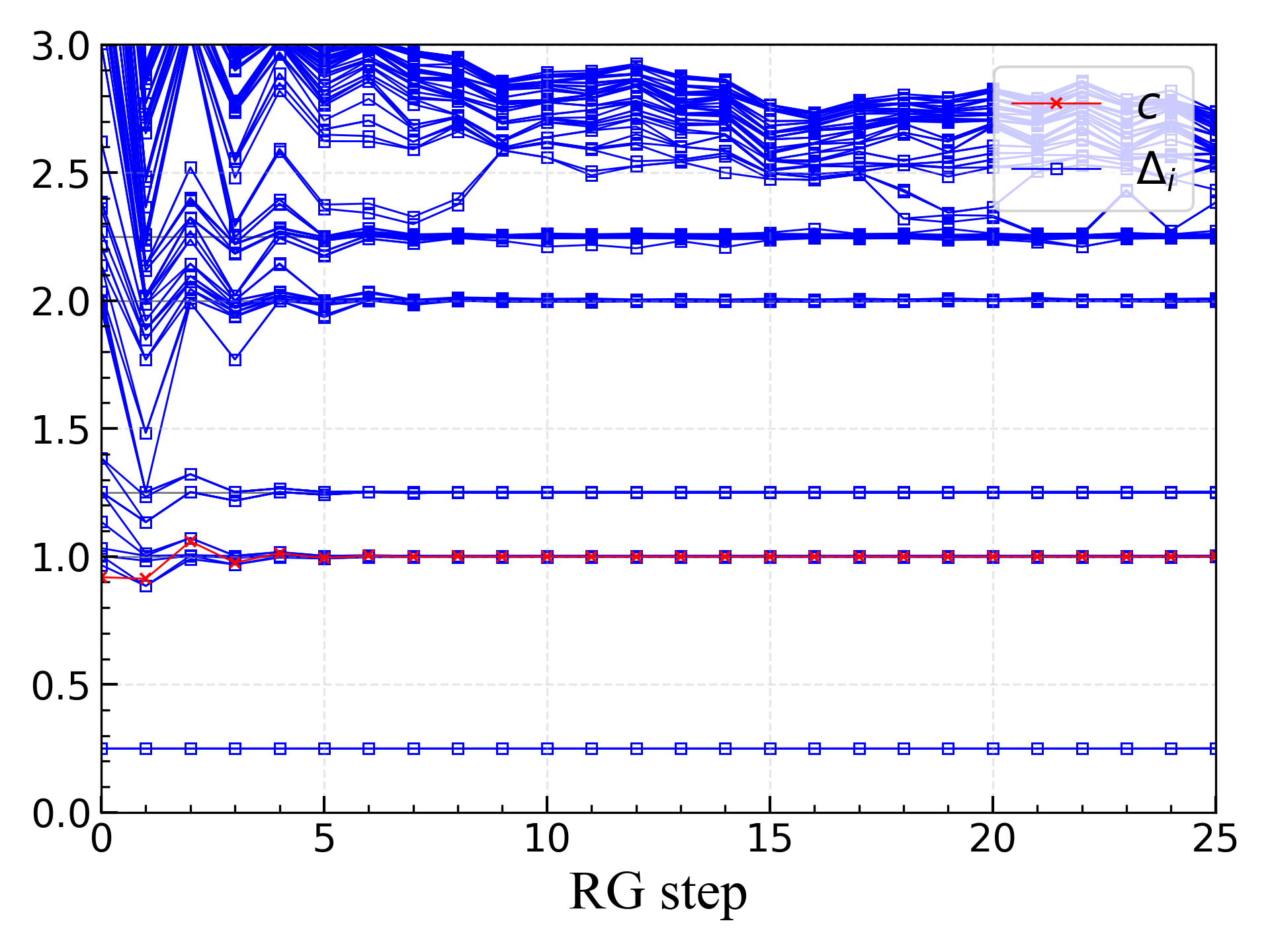}
    \caption{RG spectrum obtained from the rank-4 FP tensor at $R=1$, $\Lambda_D=4$, $w_{\max}=2$, $\chi=20$.}
    \label{fig:4pt_D_disk_1.000}
\end{figure}

\emph{Rank-4 FP tensor and marginal flow}  — 
We now construct a rank-4 tensor from a boundary four-point function, as illustrated in Fig.~\ref{fig:rank3} (b). Compared with the rank-3 construction, the rank-4 tensor has a much smaller residual marginal perturbation: in practice, retaining only the primary components already gives accurate results without imposing the MSBC. We assign Dirichlet boundary values $x_1,x_2,x_3,x_4$ to the four corners and denote by $k_{12},k_{23},k_{34},k_{41}$ the charges carried by the boundary vertex operators on the four edges. These charges satisfy the charge-conservation:
\begin{eqns}
    k_{12}+k_{23}+k_{34}+k_{41}=0.
\end{eqns}%

The tensor elements are obtained by evaluating the conformally transformed four-point
function of BCOs, as described in
the Supplemental Material. For primary fields, this gives
{\small
\begin{eqns}\label{eq:rank4T}
&\mathcal{T}^{x_1x_2x_3x_4}_{k_{12}k_{23}k_{34}k_{41}}\\
    =&|d_1|^{\frac{k_{12}^2}{2}}|d_2|^{\frac{k_{23}^2}{2}}|d_3|^{\frac{k_{34}^2}{2}}|d_4|^{\frac{k_{41}^2}{2}}|z_1-z_2|^{k_{12}\cdot k_{23}}|z_1-z_3|^{k_{12}\cdot k_{34}}\\
    \times&|z_1-z_4|^{k_{12}\cdot k_{41}}|z_2-z_3|^{k_{23}\cdot k_{34}}|z_2-z_4|^{k_{23}\cdot k_{41}}|z_3-z_4|^{k_{34}\cdot k_{41}},
\end{eqns}%
}where 
\begin{eqns}\label{eq:zd1234}
    &z_1,z_2,z_3,z_4=\left(1,-1,-(3+2\sqrt 2),3+2\sqrt 2\right),\\
    &d_1=d_2=2-\sqrt 2, \qquad d_3=d_4=2+\sqrt 2.
\end{eqns}%

We apply the same TCR analysis for the rank-4 FP tensor. The extracted central charge and conformal dimensions at $R=1$ are shown in Fig.~\ref{fig:4pt_D_disk_1.000}. Compared with the refined rank-3 FP tensor with MSBC, the rank-4 construction exhibits slightly larger residual splittings of degenerate levels. Nevertheless, the resulting RG flow remains stable and faithfully reproduces the expected conformal spectrum with
only a single set of CBCs.
Results at additional radii are provided in the Supplemental Material.



\begin{figure}[tb]
    \centering
    \includegraphics[width=.85\linewidth]{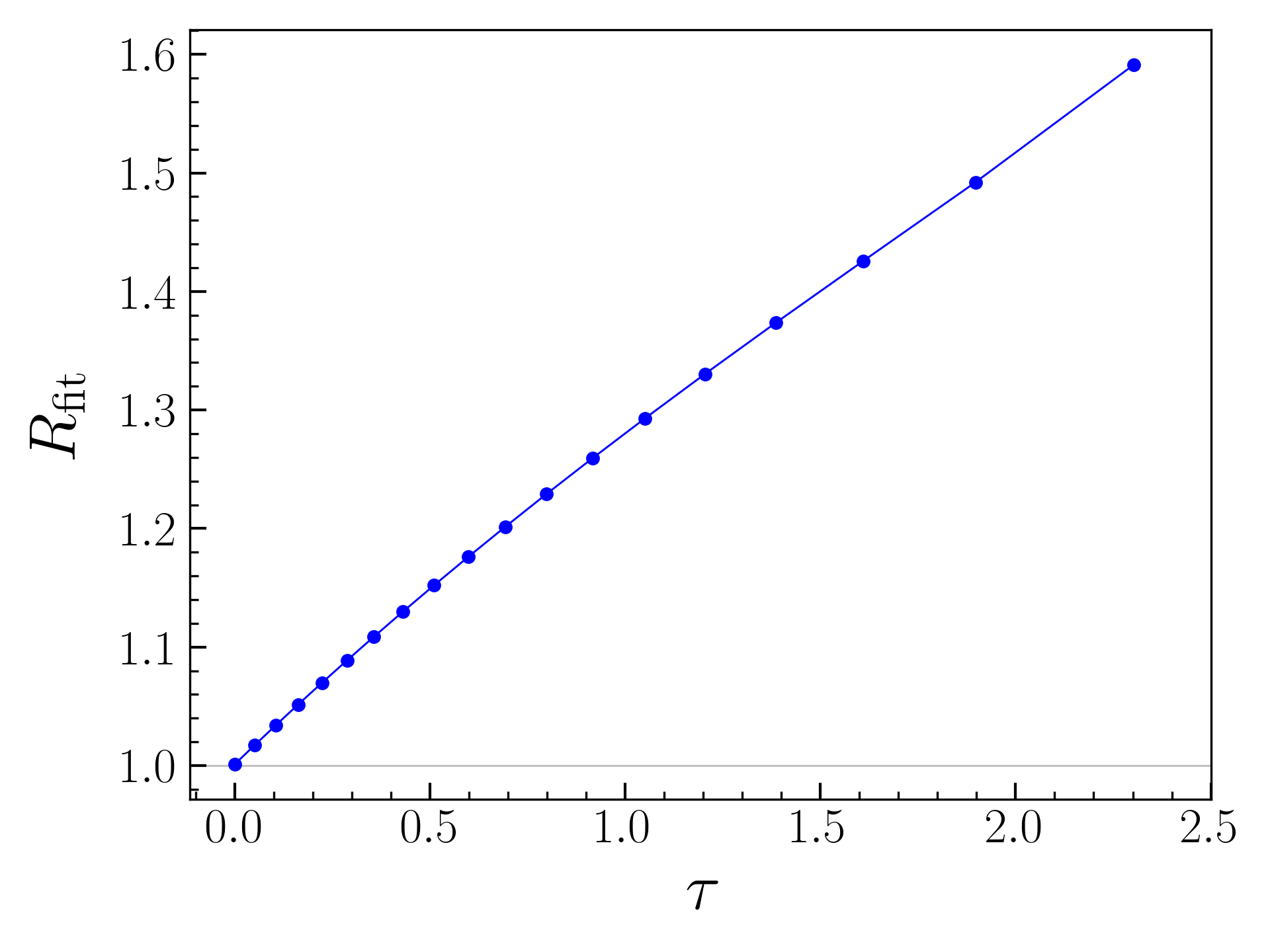}
    \caption{Change of the radius $R$ when varying $\tau$. $\tau=0$ corresponds to the radius $R=1$.}
    \label{fig:R_x_1.000}
\end{figure}

Finally, we describe a method for studying marginal flows within the FP tensor framework. Using the level-$(1,1)$ $U(1)$ descendant contribution in Eq.~\eqref{eq:DNcardy}, we can actively generate a marginal flow by modifying the geometry of the tensor.  For the rank-4 tensor shown in Fig.~\ref{fig:rank3} (b), the marginal deformation is localized at the corners of the square patch. We parametrize the corner size by $\tau$, which modifies the geometric factor in Eq.~\eqref{eq:zd1234} according to
\begin{equation}
    (d_1,d_2,d_3,d_4)\rightarrow (d_1,d_2,d_3,d_4)\cdot e^{-\tau}.
\end{equation}

We also use TCR to verify that varying $\tau$ generates an exactly marginal flow. As shown in Fig.~\ref{fig:R_x_1.000}, the family of deformed rank-4 FP tensors starting from  $R=1$ accurately reproduces the closed-string conformal spectra of compactified boson CFTs along the marginal line. Additional details on the relation between $\tau$ and the effective marginal coupling are provided in the Supplemental Material.




\emph{Conclusion and discussion}  —
We constructed FP tensors for the compactified boson CFT at generic radius, extending FP tensor constructions beyond rational CFTs. 
Although the shrinkable boundary condition is formally defined by integrating over a continuous family of CBCs, we find that the construction can be implemented to high accuracy using only a finite, discretized set of CBCs. We validated these tensors by extracting the closed-string spectrum from the transfer matrix, finding systematic improvement as more descendant states are retained. The tensors also exhibit stable RG behavior under TCR algorithm, even when restricted to the primary sector. 

We note that at finite compactification radius $R$, any grid size $\Lambda_D$ or $\Lambda_N$ satisfying Eq.~\eqref{eq:Lambdas} gives a valid discrete FP tensor. In the infinite-bond-dimension limit, all these different discretizations represent the same CFT path integral. In fact, we can also describe the continuous limit of this construction. By taking the grid size to infinity, $\Lambda\rightarrow\infty$, we can convert the finite sum over boundary conditions into an integral and gives a continuous FP tensor network, see End Matter for more details.

Our construction further allows marginal deformations to be encoded directly at the tensor level, giving a nonperturbative way to generate flows between compactified boson CFTs at different radii. This may provide a route to exploring broader CFT moduli spaces, including the circle and orbifold branches of the compactified boson~\cite{Dijkgraaf1988C1CF,Ginsparg1988CuriositiesAC}. Finally, it would be interesting to construct the ground-state wave function directly from the FP tensor. We leave these directions to the future work. 

\emph{Acknowledgement}  —
We acknowledge helpful discussion with Janet Ling-Yan Hung, Shu-Heng Shao, Yi-Kun Jiang and Chenqi Meng. This work is supported by funding from Hong Kong’s Research Grants Council (RFS2324-4S02, CRF C7015-24G, CRS HKU701/24).

\bibliographystyle{apsrev4-1}
\bibliography{Free_boson/ref}

\clearpage
\section*{End Matter}
The end matter provides further details on: (A) the derivation of Eq.~\eqref{eq:DNcardy}, and (B) the construction of FP tensor with MSBC.

\setcounter{secnumdepth}{2}
\renewcommand{\thesubsection}{\Alph{subsection}}
\setcounter{subsection}{0}

\subsection{Shrinkable boundary condition and continous FP tensor}
In the RCFT
construction, the weights $\omega_a$ in Eq.~\eqref{eq:partition} are chosen to impose the shrinkable boundary
condition \footnote{This condition is also called the entanglement-brane condition in
Ref.~\cite{PhysRevD.104.026012} and the cloaking boundary condition in
Ref.~\cite{Brehm:2021wev}.}, i.e., the weighted superposition of Cardy states projects onto the
vacuum Ishibashi state,
\begin{eqns}\label{eq:shrink}
    \sum_a \omega_a\ket{a}_c=\ket{0}\rangle.
\end{eqns}%
For an RCFT, this is a finite sum over CBCs. For the compactified boson, a simple example of an irrational
CFT, the corresponding condition becomes an integral over a continuous parameter~\cite{PhysRevD.104.026012}. For Dirichlet boundary, the condition Eq.~\eqref{eq:shrink} becomes,
\begin{eqns}\label{eq:shrinkable}
    \frac{1}{2\pi R}\int_0^{2\pi R} dx\  \ket{D,x}_c&=(2R)^{-1/2}\ket{0,0}\rangle_D,
\end{eqns}%
while for Neumann boundary, the condition is
\begin{eqns}
        \frac{R}{\pi}\int_0^{\frac{\pi}{R}} d\theta \ \ket{N,\theta}_c&=R^{1/2}\ket{0,0}\rangle_N. 
\end{eqns}%
This indicates the choice of weights to be  $\omega_D=(\pi\sqrt{2R})^{-1}$ for all $x$ and $\omega_N=\sqrt{R}/\pi$ for all $\theta$. In the above equations,  the Ishibashi states are defined with respect to the $U(1)$ current algebra,
$[J_r,J_s]=r\delta_{r+s,0}$, as
\begin{eqns}
\ket{k_L,k_R}\rangle_D &:=
\exp\left(\sum_{r=1}^{\infty}\frac{1}{r}\,J_{-r}\bar J_{-r}\right)\ket{k_L,k_R},
\\
\ket{k_L,k_R}\rangle_N &:=
\exp\left(-\sum_{r=1}^{\infty}\frac{1}{r}\,J_{-r}\bar J_{-r}\right)\ket{k_L,k_R}.
\end{eqns}%
For a detailed discussion of Cardy and Ishibashi states in the compactified boson CFT, see
Refs.~\cite{Recknagel_1999,Gaberdiel_2001,Gaberdiel_2002}.

To construct a valid FP tensor, it is sufficient to work entirely within either the Dirichlet or the
Neumann family of boundary conditions. The two resulting tensor networks are related by T-duality. Both yield equivalent CFT path integrals through the following tensor network contraction involving continuous integral over the boundary labels, 
\begin{align}
Z_{M} =  \sum_{\{ (i,I)\} } \prod_v \int da\ \omega_{a}    
\prod_{\triangle}\mathcal{T}^{a b c}_{(i,I) (j,J)(k,K)}. 
\end{align}

In this equation, the indices $i$ labels the  primary field through its $U(1)$ charge $k_i$, while $I$ labels the descendants. 

To obtain a discrete tensor-network representation of the Euclidean path integral, we replace these
continuous integrals by finite sums over grids of boundary conditions,
\begin{eqns}
    x_m&=\frac{m}{\Lambda_D} 2\pi R, \qquad m=0, 1,2, \cdots \Lambda_D-1\\
    \theta_l&=\frac{l}{\Lambda_N}\frac{\pi}{R}, \qquad l=0,1,2, \cdots \Lambda_N-1
\end{eqns}%
We denote the corresponding Cardy states by $\ket{D,x_m}_c$ and $\ket{N,\theta_l}_c$. The equal-weight superposition over either grid selects the vacuum Ishibashi state together with a tower of higher-charge primaries:
\begin{eqns}\label{EM:DNcardy}
    &\frac{1}{\Lambda_D}\sum_{m=0}^{\Lambda_D-1} \ket{D,x_m}_c\\
    =&(2R)^{-1/2}\sum_{m\in \mathbb{Z}}\ket{\frac{m\Lambda_D}{2R},\frac{m\Lambda_D}{2R}}\rangle_D\\
    =&(2R)^{-1/2}\left(\ket{0}+J_{-1}\bar J_{-1}\ket{0}+\ket{\frac{\Lambda_D}{2R},\frac{\Lambda_D}{2R}}+\cdots\right),\\
    &\frac{1}{\Lambda_N}\sum_{l=0}^{\Lambda_N-1}\ket{N, \theta_l}_c\\
    =&R^{1/2}\sum_{l\in \mathbb{Z}}\ket{l\Lambda_N R,-l\Lambda_N R}\rangle_N\\
    =&R^{1/2}\left(\ket{0}-J_{-1}\bar J_{-1}\ket{0}+\ket{\Lambda_N R,-\Lambda_N R}+\cdots\right).
\end{eqns}%
The leading corrections within the vacuum module are the level-$(1,1)$ descendants
$J_{-1}\bar J_{-1}\ket{0}$, with conformal weights $(h,\bar h)=(1,1)$. These correspond to marginal
deformations of the vacuum state. The remaining leading corrections are higher-charge primaries with
scaling dimensions
\begin{eqns}
    \Delta_{\ket{\frac{\Lambda_D}{2R},\frac{\Lambda_D}{2R}}}=(\frac{\Lambda_D}{2R})^2,\\
    \Delta_{\ket{\Lambda_N R,-\Lambda_N R}}=(\Lambda_N R)^2.
\end{eqns}%
To make these corrections irrelevant, we require $\Delta>2$ for both operators. This gives the
cutoff conditions Eq. (\ref{eq:Lambdas}).
Any $\Lambda_D$ or $\Lambda_N$ satisfying this condition defines a valid discrete representation of the FP tensor. 

The decompactification limit $R\rightarrow\infty$ is qualitatively different. In this limit, the $U(1)$ charge spectrum itself becomes continuous, so the tensor network is continuous not only in its boundary-condition labels but also in its tensor-leg labels. For example, choosing Neumann boundary conditions, we may label each open-string primary by a continuous momentum $p$. The sum over charge sectors is then replaced by an integral, and the partition function takes the schematic form
\begin{align}
Z_{M} =  \sum_{\{ I\} } \prod_e \int dp_i 
\prod_{\triangle}\mathcal{T}_{(i,I) (j,J)(k,K)}. 
\end{align}

The FP tensor network becomes intrinsically continuous in the this limit, requiring new tools for contraction, renormalization, and extraction of closed-string spectra. This may also connect spacetime FP tensors to continuous tensor-network states for quantum field theories~\cite{Verstraete_2010, Jennings_2015, PhysRevX.9.021040, PhysRevResearch.3.023059, PhysRevD.105.045016}.

\subsection{Refined FP tensor with MSBC}
The Cardy state associated with the MSBC is given by the following weighted linear combination of the discretized Dirichlet and Neumann Cardy states:
\begin{eqns}
&\ket{\mathcal{B}_{\rm mix}}\\
:=
    &\frac{1}{2}\left(\frac{(2R)^{1/2}}{\Lambda_D}\sum_{m=0}^{\Lambda_D-1} \ket{D,x_m}_c+\frac{1}{R^{1/2}\Lambda_N}\sum_{l=0}^{\Lambda_N-1}\ket{N,\theta_l}_c\right).
\end{eqns}%
Using the expansions in Eq.~\eqref{EM:DNcardy}, we obtain
\begin{eqns}
&\ket{\mathcal{B}_{\rm mix}}\\
=&\sum_{m\in \mathbb{Z}}\ket{\frac{m\Lambda_D}{2R},\frac{m\Lambda_D}{2R}}\rangle_D+\sum_{l\in \mathbb{Z}}\ket{m\Lambda_N R,-m\Lambda_N R}\rangle_N\\
    =&\ket{0}+\frac{1}{2}\ket{\frac{\Lambda_D}{2R},\frac{\Lambda_D}{2R}}+\frac{1}{2}\ket{\Lambda_N R,-\Lambda_N R}+\frac{1}{2}J_{-1}^2\bar J_{-1}^2\ket{0}\\
    &+\frac{1}{2}\ket{\frac{-\Lambda_D}{2R},\frac{-\Lambda_D}{2R}}+\frac{1}{2}\ket{-\Lambda_N R,+\Lambda_N R} +\cdots
\end{eqns}%
with the level-$(1,1)$ marginal contribution $J_{-1}\bar J_{-1}\ket{0}$ in the vacuum sector canceled. 

We now construct a refined FP tensor that incorporates the MSBC. The rank-3 tensor is still defined from boundary three-point functions of BCOs, but the corner degrees of freedom $a,b,c$ are now allowed to take values in the Dirichlet or Neumann families. As a result, the refined tensor necessarily couples the Dirichlet and Neumann sectors.

When all three boundary conditions are of the same type, either $D$ or $N$, the tensor elements are still determined by the vertex-operator correlator in Eq.~\eqref{eq:VVV}. By contrast, when one boundary segment is of $D$-$N$ type, the relevant correlator involves the twist fields $\sigma$ and $\bar\sigma$, which have conformal weights
$h_\sigma=h_{\bar\sigma}=1/16$. In particular, for boundary conditions of type $N$-$N$-$D$ (parametrized by $\theta_a,\theta_b,x_c$), one has~\cite{2000NuPhB.583..381F,erler_string_2014}
\begin{eqns}\label{eq:sVs1}
    &\langle \bar\sigma(z_1)V_k^{ab}(z_2)\sigma(z_3)\rangle\\
    =&4^{-h_{ab}}e^{i (\frac{n}{R}+ \frac{\theta_b-\theta_a}{2\pi})x_c}|z_{12}|^{-h_{ab}}|z_{23}|^{-h_{ab}}|z_{31}|^{h_{ab}-1/8},
\end{eqns}%
where $h_{ab}=\frac{1}{2}k^2$ and $k=k_{ab}^{(N\!-\!N)}$ is given in Eq.~\eqref{eq:N-Nk}. Similarly, for boundary conditions of type $D$-$D$-$N$ (parameterized by $x_a,x_b,\theta_c$), one finds the following.
\begin{eqns}\label{eq:sVs2}
&\langle \sigma(z_1)V_k^{ab}(z_2)\bar\sigma(z_3)\rangle\\
    =&4^{-h_{ab}}e^{i (2wR + \frac{x_b-x_a}{2\pi})\theta_c}|z_{12}|^{-h_{ab}}|z_{23}|^{-h_{ab}}|z_{31}|^{h_{ab}-1/8},
\end{eqns}%
 where now $k=k_{ab}^{(D\!-\!D)}$ is given in Eq.~\eqref{eq:D-Dk}.

The primary components of the refined FP tensor are given by Eq.~\eqref{eq:primarytensor}, with the structure constant read from Eq.~\eqref{eq:sVs1} and Eq.~\eqref{eq:sVs2}.

\clearpage


\begin{center}
\textbf{Supplemental Material}
\end{center}




\setcounter{equation}{0}
\renewcommand{\theequation}{S\arabic{equation}}

\setcounter{figure}{0}
\renewcommand{\thefigure}{S\arabic{figure}}

\setcounter{table}{0}
\renewcommand{\thetable}{S\arabic{table}}

\section{Detailed calculation of FP tensor}\label{app:FPtensor}
\subsection{Rank-3 FP tensor}
In this section, we review the rank-3 FP tensor for an isosceles right triangle patch \cite{chengPrecisionReconstructionRational2025}. We take the length of the edge to be $l$; by scale invariance, the final tensor elements are independent of $l$, but keeping
$l$ explicit helps define the coordinates and conformal maps.

\begin{figure}[h]
    \centering
    \includegraphics[width=0.4\linewidth]{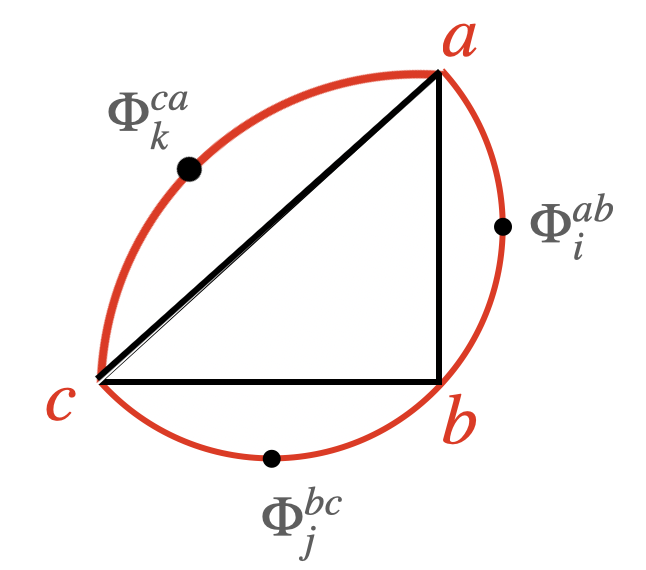}
    \caption{Rank-3 tensor as three-point function on disk.}
    \label{fig:threepoint}
\end{figure}

A rank-3 tensor element is the amplitude associated with three boundary states living on the three
edges of the triangle. Operationally, we compute this amplitude by attaching three thin boundary
segments to the triangle, as shown in Fig.~\ref{fig:threepoint}, thus obtaining a disk with three 
boundary intervals carrying CBCs $a,b,c$. By state--operator correspondence, a state $\ket{\phi_i^{ab}}$ on an edge
interpolating boundary condition $a\to b$ is in one-to-one correspondence with a  BCO $\Phi_i^{ab}$ inserted at the interface of two intervals.

\begin{figure}[h]
    \centering
    \includegraphics[width=0.3\linewidth]{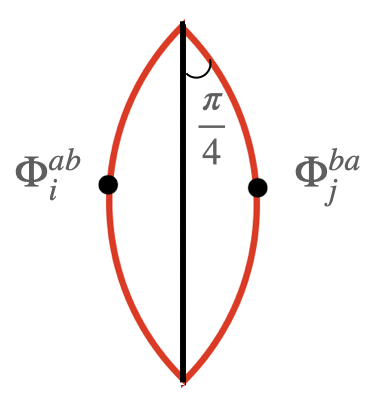}
    \caption{Two point function on nut-shape region is the inner product of states on the triangle edges.}
    \label{fig:nut}
\end{figure}

Triangles are glued along an edge by contracting the corresponding state labels with the inverse
edge metric $g^{ij}$, defined as the inverse of the inner-product matrix
\begin{eqns}
    g_{ij} = \langle \phi_i^{ab}|\phi_j^{ba} \rangle. 
\end{eqns}%
Geometrically, this inner product is computed as a two-point function on the ``nut'' region shown in
Fig.~\ref{fig:nut}:
\begin{eqns}\label{eq:edge_innerproduct}
    \langle \phi_i^{ab}|\phi_j^{ba} \rangle=\langle \Phi^{ab}_i(-x_0) \Phi^{ba}_j(x_0)\rangle_{\text{nut}}, 
\end{eqns}%
where the insertion points are located at
$x_0=\frac{\sqrt 2 -1}{2}\,l$ and $-x_0$ along the boundary.

We evaluate the nut correlator by mapping it to the upper half-plane (UHP). Let $f$ denote the
conformal map from the UHP to the nut region. Then
\begin{eqns}
    \langle \Phi^{ab}_i(-x_0) \Phi^{ba}_j(x_0)\rangle_{\text{nut}} = \langle [f^{-1}_*\Phi^{ab}_i(0)]^{\dagger}f^{-1}_*\Phi^{ba}_j(0)\rangle_{\text{UHP}}.
\end{eqns}%
The map $f$ is conveniently written as a composition
\begin{eqns}\label{eq:fz}
    &f(z) = \xi \circ \eta \circ \omega(z)\\
    &\omega(z) = \frac{1+z}{1-z}, \quad \eta(\omega)=e^{-\frac{i\pi}{4}}\omega^{1/2}, \quad \xi(\eta)=i\frac{l}{2}\frac{\eta-1}{\eta+1}.
\end{eqns}%

To simplify gluing, we choose a BCO basis in which the edge states are orthonormal. Concretely, we
write $\Phi_i^{ab}=f_*O_i^{ab}$ where $\{O_i^{ab}\}$ is a canonical basis of boundary primaries and
descendants on the UHP satisfying
\begin{eqns}\label{eq:composite}
    \langle [O_i^{ab}(0)]^{\dagger}O_j^{ba}(0)\rangle_{\text{UHP}}=\delta_{ij}.
\end{eqns}

To place the three operators at the appropriate locations on the disk boundary (Fig.~\ref{fig:threepoint}),
we introduce three local coordinate maps $f_m$ obtained from $f$ by rigid translations and rotations:
\begin{eqns}
    f_1(z) =& \left(f(z)+\frac{l}{2}\right),\\
    f_2(z) =& -i\left(f(z)+\frac{l}{2}\right),\\
    f_3(z)= &\ (i-1)f(z).
\end{eqns}%

Finally, we use an $SL(2,\mathbb{C})$ transformation $g$ to map the resulting disk configuration to a
convenient set of insertion points on the real line of the UHP:
\begin{eqns}
    g(\xi)=\left(-i\frac{\xi+\frac{\sqrt{2}l}{2}e^{i\frac{\pi}4}}{\xi-\frac{\sqrt{2}l}{2}e^{i\frac{\pi}{4}}}\right)^{4/3}.
\end{eqns}%

We define $\chi_m := g\circ f_m$ and denote the insertion points by $z_m\equiv \chi_m(0)$. With these definitions, the rank-3 tensor element is the UHP three-point function
\begin{eqns}
    T_{ijk}^{abc}=\langle \chi_{1*}O^{ab}_i(z_1)\chi_{2*}O^{bc}_j(z_2)\chi_{3*}O^{ca}_k(z_3)\rangle_{\text{UHP}},
\end{eqns}%
where $\chi_{m*}$ denotes the conformal transformation of operators under $\chi_m$. For the special case in which all three operators are primaries with weights $(h_i,h_j,h_k)$, conformal
covariance gives
\begin{eqns}
    T_{ijk}^{abc}=\frac{|\chi_1'(0)|^{h_i}|\chi_2'(0)|^{h_j}|\chi_3'(0)|^{h_k}}{|z_1-z_2|^{h_i+h_j-h_k}|z_1-z_3|^{h_i+h_k-h_j}|z_2-z_3|^{h_j+h_k-h_i}}
\end{eqns}%

Evaluating the geometric factors for the isosceles right triangle maps yields 
\begin{eqns}
    T_{h_1h_2h_3}^{abc}=C_{h_1h_2h_3}^{abc}p^{h_1+h_2}q^{h_3},
\end{eqns}%
with $p\approx0.26658$ and $q\approx0.70421$.
Note that the tensor element is independent of the edge length $l$, as expected.

\subsection{Rank-4 FP tensor}
\begin{figure}
    \centering
    \includegraphics[width=0.7\linewidth]{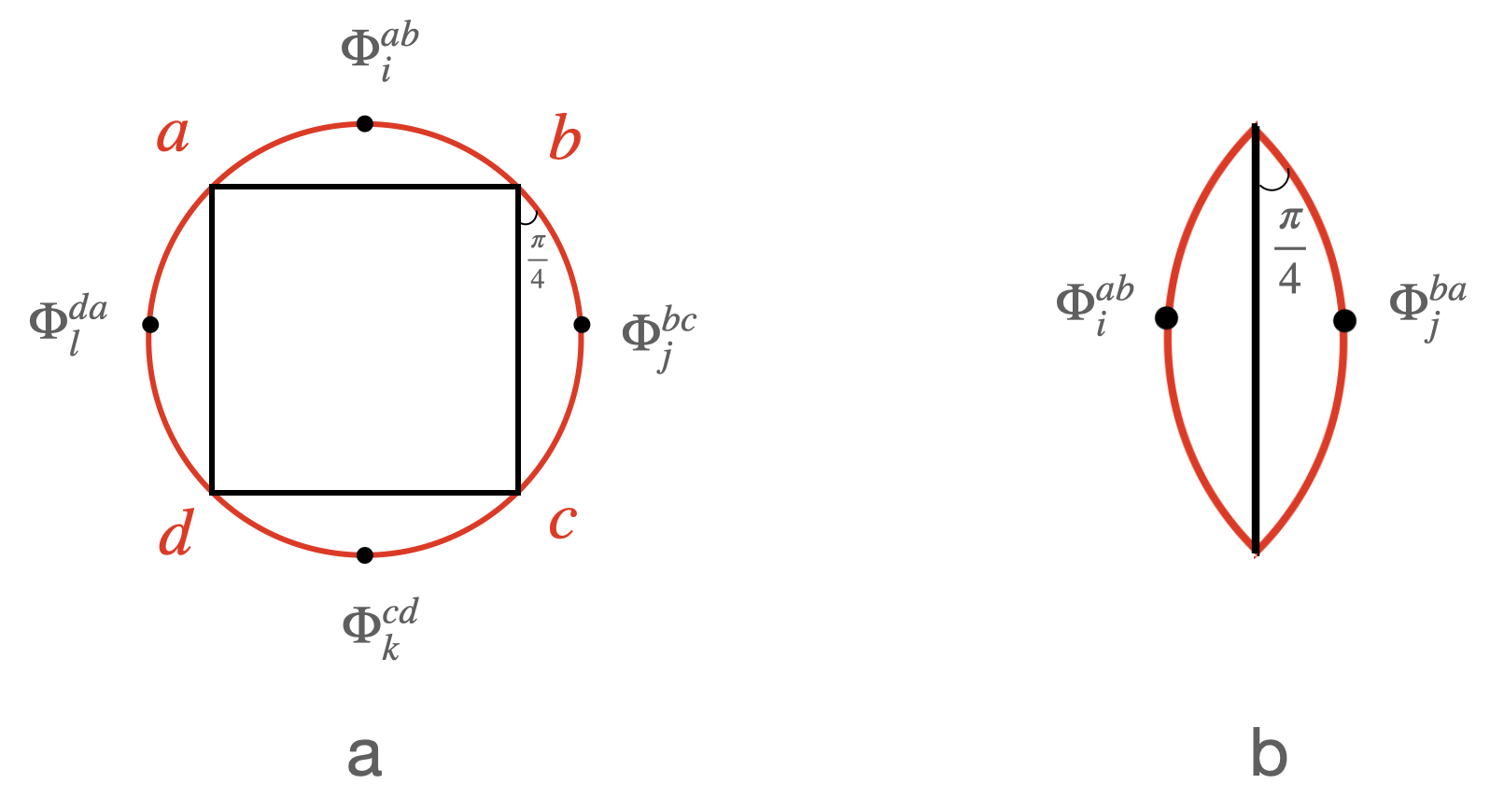}
    \caption{Rank-4 tensor as four-point function on disk.}
    \label{fig:square}
\end{figure}

The rank-4 tensor is the amplitude associated with four boundary states living on the edges of a
square patch. We take the edge length to be $l$. Similar to the rank-3 construction, a rank-4 tensor
element can be expressed as a boundary four-point function on a disk, obtained by attaching four
boundary segments to the square, as shown in Fig.~\ref{fig:square}(a). By state--operator correspondence, the
edge states are represented by BCOs $\Phi_i^{ab}$ inserted on
the disk boundary. We choose the BCO basis such that the induced edge inner product is orthonormal,
i.e., the two-point function on the corresponding ``nut'' region (Fig.~\ref{fig:square}(b)) satisfies
$\langle \Phi_i^{ab}\Phi_j^{ba}\rangle_{\rm nut}=\delta_{ij}$.

To implement this normalization, we use the same construction as in the rank-3 case.
The conformal map $f(z)$ is still given by Eq.~\eqref{eq:fz}.
We then introduce four local coordinate maps $\tilde f_m$ (one for each edge insertion), obtained
from $f(z)$ by rigid translations and $\pi/2$ rotations:
\begin{eqns}
    \tilde f_m(z)=\left(f(z)+\frac{l}{2}\right)e^{i\pi-i\frac{m \pi}{2}}, \quad m= 1,2,3,4.
\end{eqns}%

Next, we map the disk configuration to a set of insertion points on the real axis of UHP using an
$SL(2,\mathbb{C})$ transformation. A convenient choice is 
\begin{eqns}\label{eq:grank4}
    \tilde g(\xi) = \frac{\xi-\frac{\sqrt 2}{2}le^{i\pi/4}}{\xi + \frac{\sqrt 2}{2}le^{i\pi/4}}\frac{1+e^{i\pi/4}}{1-e^{i\pi/4}}. 
\end{eqns}%

In parallel with the rank-3 case, we define $\tilde\chi_m:=\tilde g\circ\tilde f_m$ and denote the
resulting insertion points by $\tilde z_m=\tilde\chi_m(0)$. The rank-4 tensor is then the UHP correlator
\begin{eqns}
    T_{ijkl}^{abcd}=\langle\tilde\chi_{1*}O_i^{ab}(\tilde z_1)\tilde\chi_{2*}O_j^{bc}(\tilde z_2)\tilde\chi_{3*}O_k^{cd}(\tilde z_3)\tilde\chi_{4*}O_l^{da}(\tilde z_4)\rangle_{\text{UHP}},
\end{eqns}%
where $\{O_i^{ab}\}$ is the canonical orthonormal BCO basis on the UHP.

For the special case in which the boundary operators are primary vertex operators
$O^{ab}_i=e^{ik_{ab}^{(i)}X}$, conformal covariance gives
\begin{eqns}\label{eq:forvertex}
T_{ijkl}^{abcd}
=&
\prod_{m=1}^{4}\left|\tilde\chi_m'(0)\right|^{\frac{k_m^2}{2}}
\prod_{1\le m<n\le 4}\left|\tilde z_{mn}\right|^{\,k_m k_n},\\
\tilde z_{mn}:=&\tilde z_m-\tilde z_n,
\end{eqns}%
where $k_m$ denotes the charge of the operator inserted at $\tilde z_m$ (determined by the adjacent boundary
conditions, e.g.\ $k_1\equiv k_{ab}^{(i)}$, $k_2\equiv k_{bc}^{(j)}$, etc.). For our choice of maps, the insertion
points and Jacobian factors evaluate to
\begin{eqns} 
\tilde z_1,\tilde z_2,\tilde z_3,\tilde z_4=&(1,-1,-(3+2\sqrt 2),3+2\sqrt 2),\\ |\tilde\chi_1'|,|\tilde\chi_2'|,|\tilde\chi_3'|,|\tilde\chi_4'|=&(2-\sqrt 2,2-\sqrt 2,2+\sqrt 2,2+\sqrt 2).
\label{eq:zchi1234}
\end{eqns}

\section{Descendants calculation}
The $U(1)$ current descendants of  a vertex operator $V_k(z)$ take the following form:
\begin{eqns}
    J_{-m_1}J_{-m_2}\cdots J_{-m_r} V_k(z).
\end{eqns}%
Note that the ordering of the current modes does not matter, since the commutator 
\begin{eqns}
    [J_{-n},J_{-m}]=(-n)\delta_{n+m,0}
\end{eqns}%
vanishes when both modes are negative. The correlation function of these descendants can be written as 
\begin{eqns}\label{eq:correlatorJV}
    \left\langle\prod_{i=1}^n[J_{-m_{i_1}}J_{-m_{i_2}}\cdots J_{-m_{i_{r}}}V_{k_i}(z_i)] \right\rangle.
\end{eqns}%

The correlator can be calculated using the integral representation
\begin{eqns}
    \left(\prod_{i=1}^n\prod_{l=1}^{r_i}\oint_{z_i} \frac{dw_{i_l}}{2\pi i} (w_{i_l}-z_i)^{-m_{i_l}}\right) \left\langle \prod_{i=1}^n\prod_{l=1}^{r_i}J(w_{i_l})V_{k_i}({z_i})\right\rangle,
\end{eqns}%
where the current--vertex correlator can be calculated by Wick contractions. The current--current and current--vertex contractions are  
\begin{eqns}
     &J(w) J(z)\sim\frac{1}{(z-w)^2},\\
     &J(w) V_k(z)\sim \frac{k}{w-z}V_k(z).
\end{eqns}%

It is convenient to introduce a generating function for the correlators 
\begin{eqns}
    \mathcal{G}(\{z_i\};\{t_{m,i}\}):=\left\langle \prod_{i=1}^n e^{\sum_{m=1}^{\infty} t_{m,i}J_{-m}}V_{k_i}(z_i)\right\rangle, 
\end{eqns}%
so that each correlator in Eq.~\eqref{eq:correlatorJV} can be obtained as 
\begin{eqns}
    \prod_{i=1}^n\prod_{l=1}^{r_i}\partial_{t_{m_{i_l},i}}\mathcal{G}(\{z_i\};\{t_{m,i}\})|_{t_{m,i}=0}.
\end{eqns}%

The generating function can also be evaluated using Wick contractions. More specifically, using the identity 
\begin{eqns}
    \langle e^{X}\rangle =\exp\left(\langle X\rangle+\frac{1}{2}\langle X^2\rangle\right),
\end{eqns}%
for a Gaussian variable $X$, we have 
\begin{eqns}
    &\mathcal{G}(\{z_i\};\{t_{m,i}\})\\
    =&\exp\left(\sum_{m,i} t_{m,i}A_{m,i}+\frac{1}{2}\sum_{m,i,n,j}t_{m,i}t_{n,j}B_{m,i;n,j}\right)\\
    &\times \prod_{i,j} (z_i-z_j)^{k_i\cdot k_j},
\end{eqns}%
with the coefficients $A_{m,i}$ and $B_{m,i;n,j}$ given by 
\begin{eqns}
    A_{m,i}=&\sum_{j\neq i} \frac{-k_j}{(z_j-z_i)^m}\\
    B_{m,i;n,j}=&(-1)^{n+1}\frac{(m+n-1)!}{(m-1)!(n-1)!}(z_j-z_i)^{-(m+n)}.
\end{eqns}

\section{The construction of transfer matrix from FP tensors}\label{app:transfer}
\begin{figure}
    \centering
    \includegraphics[width=0.4\linewidth]{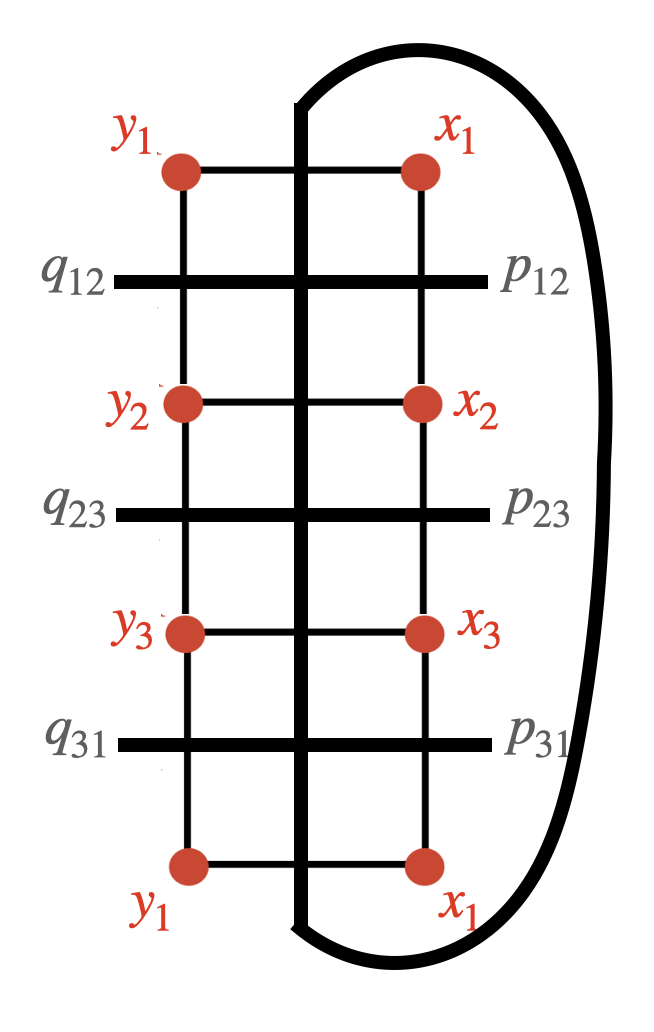}
    \caption{Transfer matrix construction.}
    \label{fig:transfer}
\end{figure}

After obtaining the rank-4 tensor, either by contracting two rank-3 tensors or by directly evaluating the boundary four-point function, we construct the transfer matrix by contracting the tensors into a cylinder, as shown in Fig.~\ref{fig:transfer}.

When two tensors are contracted along a physical leg, the corresponding $U(1)$ charges $p_1$ and $p_2$ carried by that leg must satisfy charge conservation:
\begin{align}
p_1+p_2=0 .
\end{align}
The transfer-matrix indices are therefore labeled by the Dirichlet boundary values $x_i$, together with the incoming and outgoing $U(1)$ charges $p_{ij}$ and $q_{ij}$, which satisfies 
\begin{eqns}
    \sum_{i} p_{i,i+1} = -\sum_j q_{j,j+1}.
\end{eqns}%
The boundary values $x_i$ are subject to periodic boundary conditions around the cylinder. Diagonalizing the transfer matrix then yields the closed-string spectrum. 

\begin{figure}
    \centering
    \includegraphics[width=1.0\linewidth]{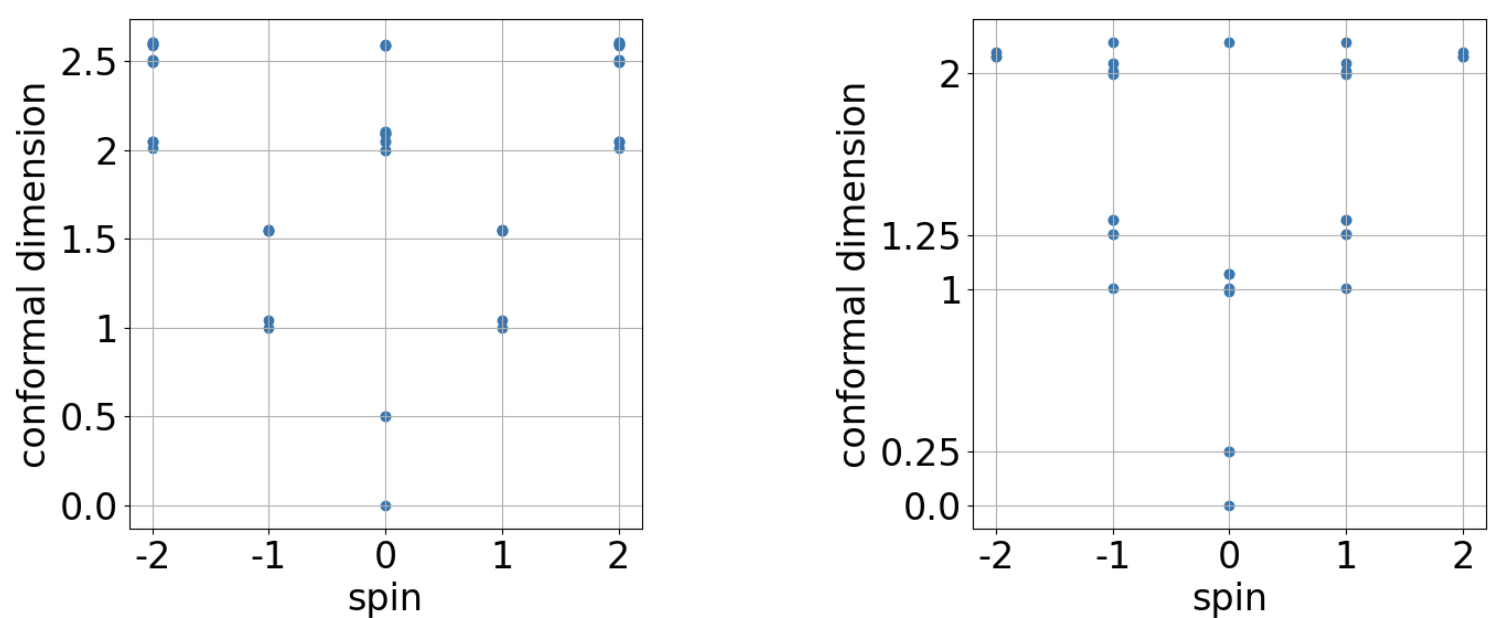}
    \caption{Conformal dimension and spin from rank-4 fixed-point tensor at the self-dual radius $R=1/\sqrt{2}$ (left) and $R=1$ (right). }
    \label{fig:data_rank4}
\end{figure}

In Fig.~\ref{fig:data_rank4} we present the closed string spectrum extracted from the transfer matrix constructed from the rank-4 tensor in Eq.~\ref{eq:forvertex}, retaining only the primary components.  

The spin quantum numbers are extracted from a twisted transfer matrix. The twist implements a one-site translation around the spatial circle of the cylinder, i.e., a cyclic permutation of the CBC and BCO labels on one side.  As an example, for the transfer matrix of length 3, let the ordinary transfer matrix be written as
\begin{eqns}
    M_{y_1q_{12}y_2q_{23}y_3q_{31}}^{x_1p_{12}x_2p_{23}x_3p_{31}},
\end{eqns}%
where the upper indices label one side of the cylinder and the lower indices label the other. The twisted transfer matrix $\tilde M$ is obtained by cyclically shifting the upper indices,
\begin{eqns}
    \tilde{M}_{y_1q_{12}y_2q_{23}y_3q_{31}}^{x_1p_{12}x_2p_{23}x_3p_{31}}:=M_{y_1q_{12}y_2q_{23}y_3q_{31}}^{x_2p_{23}x_3p_{31}x_1p_{12}}.
\end{eqns}

Then diagonalizing the twisted transfer matrix produce the spectrum: 
\begin{eqns}
    \lambda_n(j)= e^{\frac{2\pi}{n}(\frac{c}{12}-\Delta_j-is_j)}.
\end{eqns}

Here $n$ denotes the cylinder length. The conformal dimension is extracted from the real part of the exponent, while the conformal spin is extracted from its imaginary part.

\section{Tuning on the marginal deformation}

\begin{figure}
    \centering
    \includegraphics[width=0.7\linewidth]{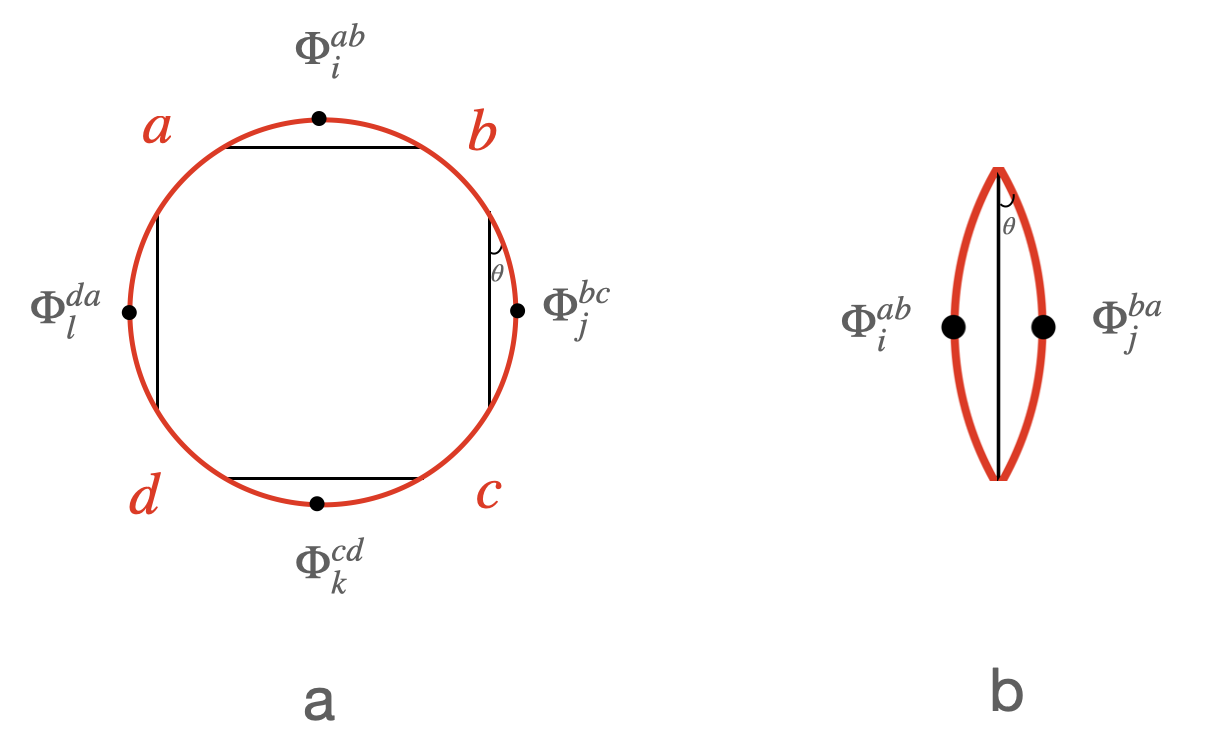}
    \caption{Rank-4 tensor geometry with corners opened into finite boundary intervals.}
    \label{fig:square_holes}
\end{figure}

In this section, we deform the rank-4 tensor geometry by rounding each corner to a finite size segment, as illustrated in Fig.~\ref{fig:square_holes}(a). Specifically, the square patch is replaced by a ``rounded square'', whose corners are opened into short boundary segments that can support
CBCs.

\begin{figure}
    \centering
    \includegraphics[width=0.3\linewidth]{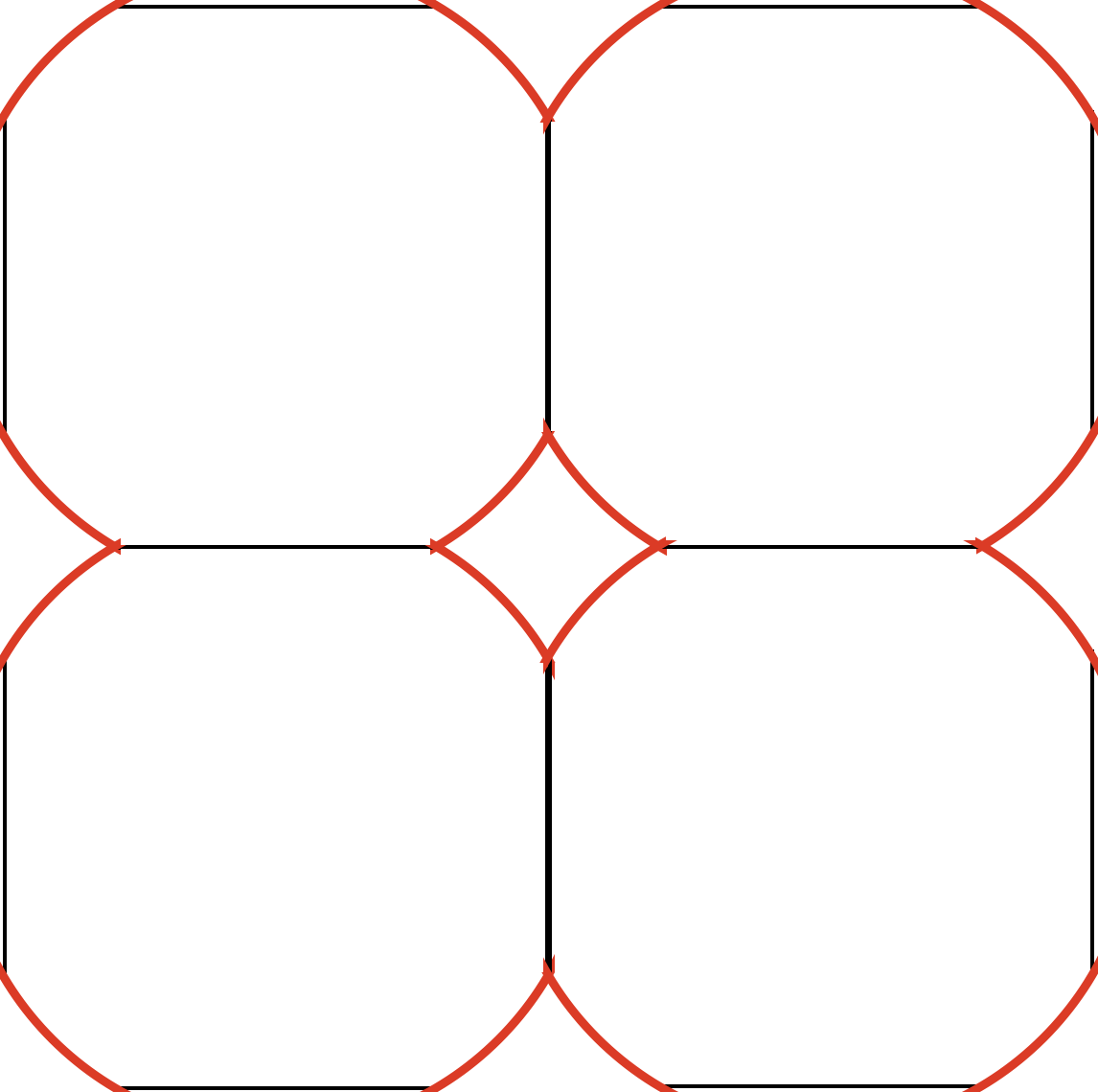}
    \caption{Rounded-square tessellation of a surface. The unglued corner intervals become holes that
    carry conformal boundary conditions. }
    \label{fig:surface_holes}
\end{figure}

When these modified rank-4 tensors are contracted into a two-dimensional network, representing the path integral on a surface, the opened corners form holes on the surface, as
shown in Fig.~\ref{fig:surface_holes}. Related geometries have also been considered in
Refs.~\cite{Brehm:2021wev,Brehm:2024zun}, although with different choices of conformal maps.  

By applying an appropriate weighted sum over the corner CBCs  on each
rank-4 tensor, we impose the shrinkable boundary condition on every hole (see
Eq.~\eqref{eq:DNcardy}). The associated boundary state is therefore the vacuum Ishibashi state of the
$U(1)$ current algebra,
\begin{eqns}
    \ket{0,0}\rangle_D+\cdots=(2R)^{-1/2}\left(\ket{0}+J_{-1}\bar J_{-1}\ket{0}+\cdots\right),
\end{eqns}%
where the ellipsis denotes irrelevant contributions.  The finite size of each hole corresponds to an effective Euclidean time evolution in the closed-string
channel. Denoting the corresponding modulus by $r$, the boundary state is dressed by
$e^{-|\log r| H_{\rm closed}}$:
\begin{eqns}
    e^{-|\log r| H_{\text{closed}}}\ket{0,0}\rangle_D = \ket{0}+e^{-2|\log r|}J_{-1}\bar J_{-1}\ket{0}+\cdots.
\end{eqns}%
Thus, at finite $r$, the marginal $(1,1)$ component survives with a tunable coefficient. In this
sense, $r$ serves as a control parameter for an effective marginal deformation of the compactified boson
CFT realized by the tensor network.

At the level of a single rank-4 tensor, the modulus $r$ can be tuned by varying the wedge angle
$\theta$,  as shown in Fig.~\ref{fig:square_holes}. This dependence is fully
encoded in the conformal maps defining the tensor. In
particular, we introduce the $\theta$-dependent map
\begin{eqns}
    f_{\theta}(z):=&\sqrt 2\sin\theta\ \xi\circ {\eta}_{\theta}\circ \omega(z),\\
    \eta_{\theta}(z):=&e^{-i\theta}\omega^{2\theta/\pi}.
\end{eqns}%
These maps preserve the orthonormality condition defined by the two-point function on the nut region shown in Fig.~\ref{fig:square_holes}(b). The function $\xi$ and $\omega$ are given in Eq.~\eqref{eq:composite}. 

Then we introduce four $\theta$-dependent local coordinate maps:
\begin{eqns}
     f^{(\theta)}_m=( f_{\theta}(z)+\frac{\sqrt 2}{2}l\cos(\theta))e^{i\pi-\frac{im\pi}{2}},
\end{eqns}%
and define $\chi_m^{(\theta)}:=\tilde g\circ f_m^{(\theta)}$, where $\tilde g$ is the same
$SL(2,\mathbb{C})$ map used in the undeformed rank-4 tensor construction (Eq.~\eqref{eq:grank4}). The resulting $\theta$-dependent rank-4 tensor is
{\small
\begin{eqns}
    [T_{\theta}]_{ijkl}^{abcd}=\langle\chi_{1*}^{(\theta)}O_i^{ab}(\tilde z_1)\chi_{2*}^{(\theta)}O_j^{bc}(\tilde z_2)\chi_{3*}^{(\theta)}O_k^{cd}(\tilde z_3)\chi_{4*}^{(\theta)}O_l^{da}(\tilde z_4)\rangle_{\text{UHP}}.
\end{eqns}
}

For primary vertex operators, the correlator takes the same form as in Eq.~\ref{eq:forvertex}
with the set of coordinates $z_m$ being 
\begin{eqns}
    \tilde z_1,\tilde z_2,\tilde z_3,\tilde z_4=\left(1,-1,-(3+2\sqrt 2),3+2\sqrt 2\right),
\end{eqns}
and $\theta$-dependent Jacobians
\begin{eqns}
    [\chi^{(\theta)}_1]'(0)=[\chi^{(\theta)}_2]'(0)=&-\frac{4 \sqrt{2} \theta  \tan \left(\frac{\theta }{2}\right)}{\pi }\\
    [\chi^{(\theta)}_3]'(0)=[\chi^{(\theta)}_4]'(0)=&-\frac{4 \left(3 \sqrt{2}+4\right) \theta  \tan \left(\frac{\theta }{2}\right)}{\pi }.
\end{eqns}%
When $\theta=\pi/4$, the construction reduces to the undeformed rank-4 tensor (cf. Eq.~\eqref{eq:forvertex}), corresponding to $r=0$. Reducing $\theta$ away from $\pi/4$ corresponds to increasing $r$, which induces a rescaling
\begin{equation}  (|\tilde\chi_1'|,|\tilde\chi_2'|,|\tilde\chi_3'|,|\tilde\chi_4'|)\rightarrow (|\tilde\chi_1'|,|\tilde\chi_2'|,|\tilde\chi_3'|,|\tilde\chi_4'|)\cdot \beta(\theta)
\end{equation}
with $0\leq \beta(\theta)\leq1$, where $|\tilde\chi_i'|$ are given by Eq.~\eqref{eq:zchi1234}.

After diagonalizing the transfer matrix, the compactification radius can be extracted by fitting the two lowest numerically obtained scaling dimensions, $\Delta_1$ and $\Delta_2$. By T-duality $R\leftrightarrow 1/2R$, we restrict to $R>1/\sqrt{2}$, so the lowest two scaling dimensions equal $1/4R^2$. Then the fitted radius is obtained by minimizing the cost function

\begin{equation}\label{eq:Rfit}
    f(R)=\left(\frac{1}{4R^2}-\Delta_{1}\right)^2+\left(\frac{1}{4R^2}-\Delta_{2}\right)^2.
\end{equation}

\section{Numerical verification by tensor complex renormalization}
In this section, we provide further numerical evidence that the FP tensor proposed in the main text is the true fixed-point tensor of the compactified boson CFT by applying the tensor complex renormalization (TCR) algorithm based on Loop-TNR adapted to the present FP-tensor construction \cite{yangLoopOptimizationTensor2017,bao2025tensorcomplexrenormalizationgeneralized}.

\begin{figure}[h]
    \centering
    \includegraphics[width=.9\linewidth]{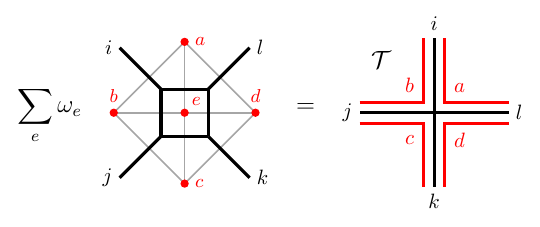}
    \caption{The initial rank-4 tensor $\mathcal{T}$ built from four rank-3 tensors.}
    \label{fig:initial_rank4_tensor}
\end{figure}

We carry out TCR on a square lattice, starting from a rank-4 tensor $\mathcal{T}$ either computed directly in the main text or built by gluing four rank-3 tensors in the shape of an isosceles right triangle, as shown in Fig.~\ref{fig:initial_rank4_tensor}.

In a rank-3 tensor, if the CBCs at the two ends of an edge are of the same type (say, the state is in the $D$-$D$ sector), the truncated Hilbert space on the edge $be$ in triangle $bea$ is labeled by

\begin{equation}\label{eq:kbe}
    k_{be}=2\left(\frac{x_b-x_e}{2\pi}+w_{be}R\right),
\end{equation}
where we adopt the convention that the boundary labels are ordered counterclockwise $b\rightarrow e\rightarrow a$ and $w_{be}\in\{-w_{\max},\dots,+w_{\max}\}$. When two rank-3 tensors are contracted (i.e., when two edges are glued together), the states on the shared edge are paired through the edge inner product \eqref{eq:edge_innerproduct}. For the compactified boson, a state carrying charge $k$ is therefore paired with a state carrying charge $-k$.
For the contraction shown on the left-hand side of Fig.~\ref{fig:initial_rank4_tensor}, if the boundary labels for triangles all go counterclockwise, then the first term in Eq.~\eqref{eq:kbe} for $k_{eb}$ in triangle $ceb$ is already reversed compared to $k_{be}$ in triangle $bea$. To make $k_{eb}=-k_{be}$, the second term $w_{eb}$ in triangle $ceb$ must also be reversed. In practice, it is convenient to arrange $w$'s in one triangle $bea$ in such a way that on edge $be$ and $ab$ they are in order $w_{be}=-w_{\max},\dots,+w_{\max}$, while on edge $ea$ it is in reverse order $w_{ea}=+w_{\max},\dots,-w_{\max}$. Then if all other three triangles are obtained by rotating the triangle $bea$, each momentum $k$ is automatically paired with $-k$ in the contraction inside the rank-4 tensor $\mathcal{T}$.

For the same reason, extra care is required in the initial contraction of the rank-4 tensors $\mathcal{T}$. In the initial coarse-graining step in the TCR algorithm, the initial leg $i(j)$ is contracted with the initial leg $k(l)$. Since their boundary labels are already oriented in opposite directions, a convenient setup is that we arrange all $w_i,w_j$ in order $-w_{\max},\dots,+w_{\max}$ while we arrange all $w_k,w_l$ in reverse order $+w_{\max},\dots,-w_{\max}$ in the initial rank-4 tensor $\mathcal{T}$ (see \figref{fig:initial_rank4_tensor}). Then we can perform the usual tensor contractions in the initial coarse-graining step (we do not perform loop optimization in the first step of TCR). In later steps, all tensor legs are generated by SVD,
so this subtlety disappears.

For the refined FP tensor with MSBCs, after gluing four rank-3 tensors, the same index-reversal procedure can be applied to the entire rank-4 tensor $\mathcal{T}$, since this setup also applies to the $N$-$N$ sector. For the mixed $D$-$N$ and $N$-$D$ sector, this procedure has no effect, since the corresponding Hilbert space initially contains only a single component, namely the twist field.

\begin{figure}
\centering
\begin{subfigure}{\linewidth}
    \centering
    \includegraphics[width=.85\linewidth]{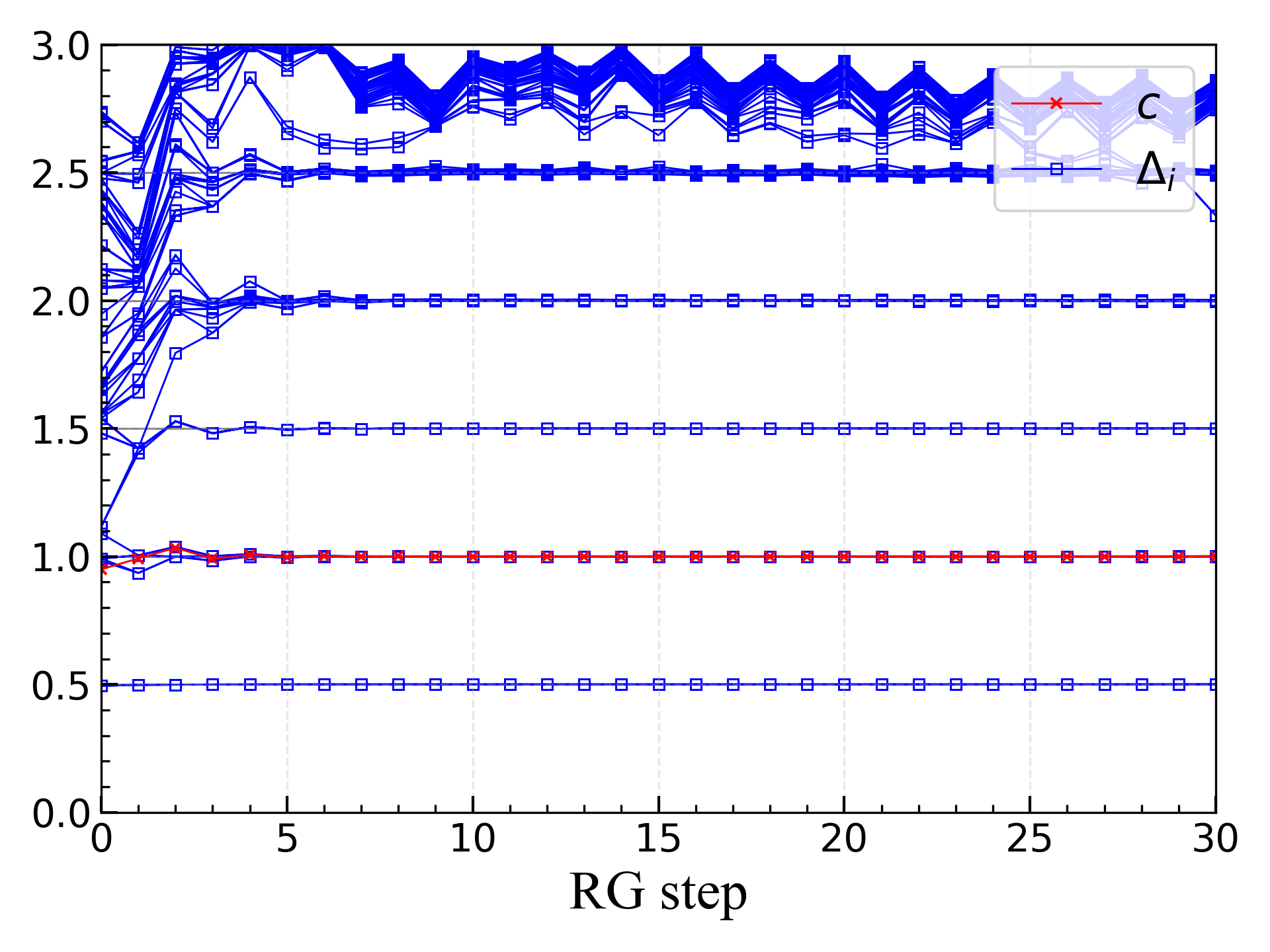}
    \caption{$R=1/\sqrt{2}$ (self-dual), $\Lambda_D=3$, $\Lambda_N=3$.}
    \label{fig:3pt_DN_disk_0.707}
\end{subfigure}
\begin{subfigure}{\linewidth}
    \centering
    \includegraphics[width=.85\linewidth]{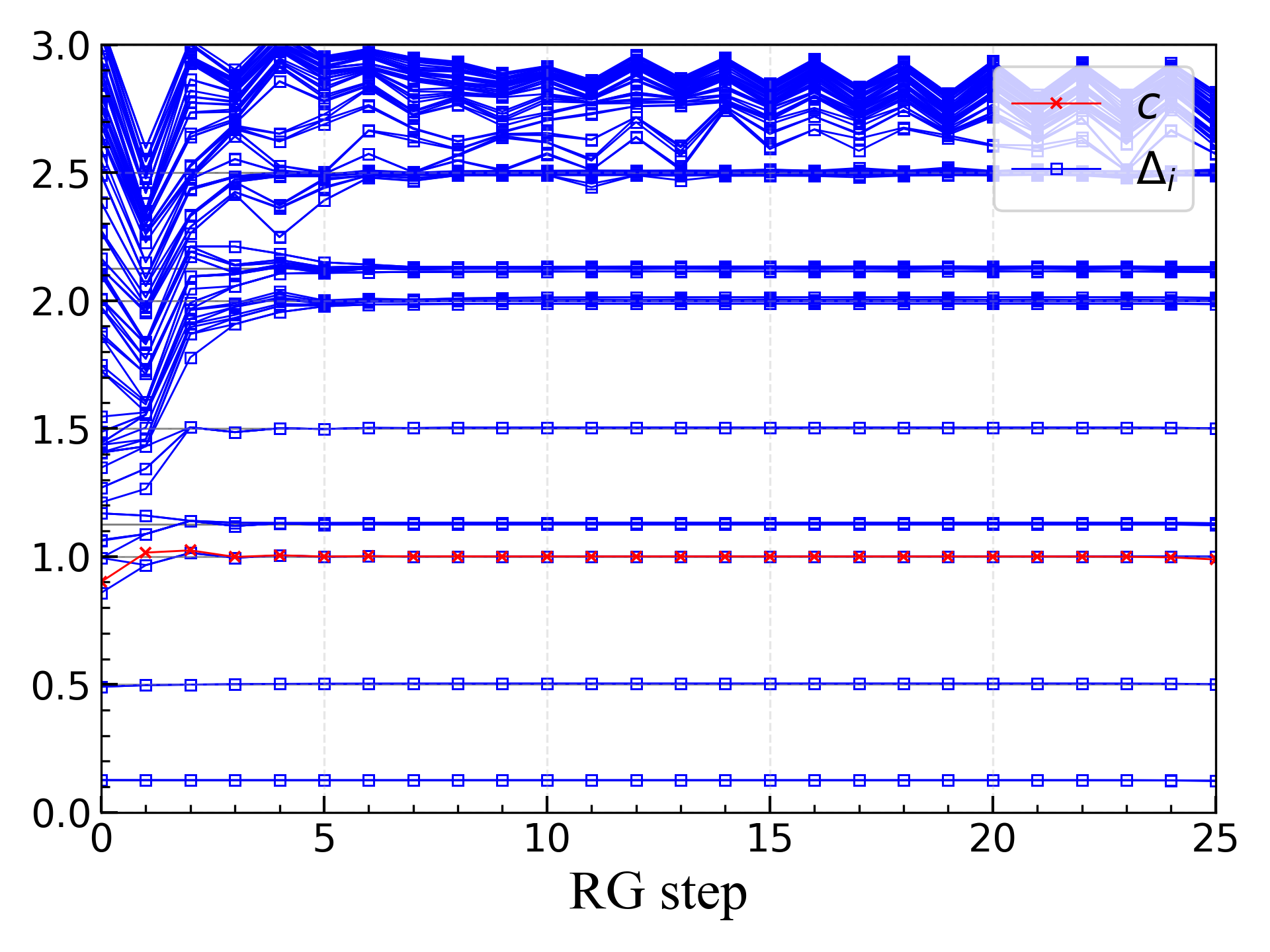}
    \caption{$R=\sqrt{2}$, $\Lambda_D=5$, $\Lambda_N=2$.}
    \label{fig:3pt_DN_disk_1.414}
\end{subfigure}
\caption{RG spectrum from the refined rank-3 FP tensor, $w_{\max}=n_{\max}=2$, $\chi=20$.}
\label{fig:3pt_DN_disk_x2}
\end{figure}

\begin{table}
\caption{
Central charge and low-lying scaling dimensions of the compactified boson CFT, initialized with the refined rank-3 FP tensor.
}
\label{tab:3pt_DN_disk_x2}
\centering
\begin{subtable}{\linewidth}
    \caption{
    $R=1/\sqrt{2}$.
    }
    \label{tab:3pt_DN_disk_0.707}
    \centering
    \begin{tabular*}{.85\columnwidth}{@{\hspace{8pt}} @{\extracolsep{\fill}} c c c @{\hspace{4pt}}}
    \hline\hline
    \multicolumn{3}{c}{RG steps 10-25: $\quad c_{\rm numerical}=1.0000(1)$} \\
    \hline\hline
    $\Delta_{\rm exact}$ & degeneracy & $\Delta_{\rm numerical}$ \\
    \hline
    0.5 & 4  & 0.5000(1) \\
    1.0 & 6  & 1.0000(2) \\
    1.5 & 8 & 1.5000(3) \\
    2.0 & 17 & 2.000(4) \\
    2.5 & 28 & 2.50(3) \\
    \hline\hline
    \end{tabular*}
\end{subtable}
\par\medskip
\begin{subtable}{\linewidth}
    \caption{
    $R=\sqrt{2}$.
    }
    \label{tab:3pt_DN_disk_1.414}
    \centering
    \begin{tabular*}{.85\columnwidth}{@{\hspace{8pt}} @{\extracolsep{\fill}} c c c @{\hspace{4pt}}}
    \hline\hline
    \multicolumn{3}{c}{RG steps 10-20: $\quad c_{\rm numerical}=1.00002(7)$} \\
    \hline\hline
    $\Delta_{\rm exact}$ & degeneracy & $\Delta_{\rm numerical}$ \\
    \hline
    0.125 & 2  & 0.1258(1) \\
    0.5 & 2  & 0.50309(4) \\
    1.0 & 2  & 1.00001(8) \\
    1.125 & 6  & 1.128(4) \\
    1.5 & 4 & 1.5030(2) \\
    2.0 & 9 & 2.00(1) \\
    2.125 & 18 & 2.12(1) \\
    \hline\hline
    \end{tabular*}
\end{subtable}
\end{table}

\subsection{RG from the refined rank-3 FP tensors}

We first verify the validity of the refined rank-3 FP tensor with MSBCs by showing that it reproduces the correct spectrum and generates stable RG flows.
We initialize the rank-4 tensor $\mathcal{T}$ by contracting four refined rank-3 FP tensors with $\Lambda_D$ Dirichlet CBCs and $\Lambda_N$ Neumann CBCs. They are chosen such that
\begin{equation}\label{eq:app_lambdaDN}
    \Lambda_D>2\sqrt{2}R\quad\text{and}\quad\Lambda_N>\frac{\sqrt{2}}{R},
\end{equation}
as explained in the main text. In addition to the $R=1$ (the free Dirac fermion point) case shown in the main text, the spectra for (1) $R=1/\sqrt{2}$ (the self-dual $SU(2)_1$ WZW point) and (2) $R=\sqrt{2}$ (the KT point) along the RG flow are shown in Fig.~\ref{fig:3pt_DN_disk_x2} (here $\chi$ denotes the TCR cutoff bond dimension). The numerical estimates for the central charge and scaling dimensions are listed in \tabref{tab:3pt_DN_disk_x2}. They are obtained by averaging over the central region of the stable RG plateau. The quoted uncertainties are estimated from the maximal deviation within the averaging window, since results from different RG steps are not statistically independent. Specifically, let $s_1$ and $s_2$ denote the first and last RG steps in the averaging window, and define $N_s=s_2-s_1+1$. The estimate of the central charge is
\begin{equation}
    c_{\text{num}}=\frac{1}{N_s}\sum_{i=s_1}^{s_2}c_i,\quad \delta c=\max_{s_1\leq i\leq s_2}|{c_i-c_{\text{num}}}|.
\end{equation}
For scaling dimensions, let $d$ denote the degeneracy (multiplicity) of the corresponding conformal level, including both primary and descendant states. The estimate is obtained by averaging over both the RG window and the $d$ states belonging to that level:
\begin{equation}
    \Delta_{\text{num}}=\frac{1}{dN_s}\sum_{i=s_1}^{s_2}\sum_{a=1}^d \Delta_{i,a},\quad \delta \Delta=\max_{\substack{s_1\leq i\leq s_2\\1\leq a\leq d}}|{\Delta_{i,a}-\Delta_{\text{num}}}|.
\end{equation}

We find that the refined rank-3 FP tensors generate stable RG flows under TCR. The resulting spectra agree with the exact values to high accuracy and exhibit the correct degeneracy structure.

\subsection{RG from the rank-4 FP tensors}

\begin{figure}[tb]
\centering
\begin{subfigure}{\linewidth}
    \centering
    \includegraphics[width=.85\linewidth]{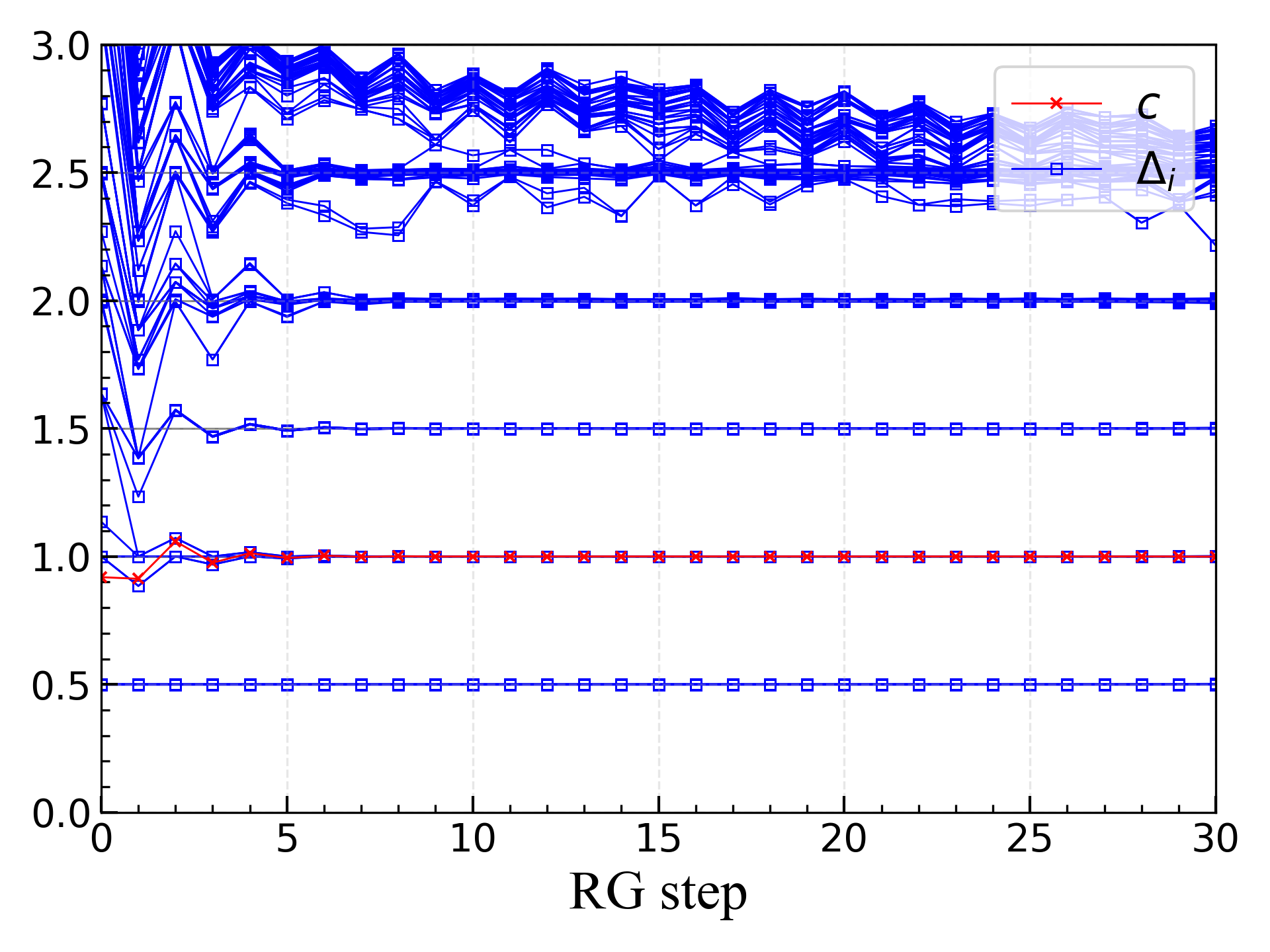}
    \caption{$R=1/\sqrt{2}$ (self-dual), $\Lambda_D=4$.}
    \label{fig:4pt_D_disk_0.707}
\end{subfigure}
\begin{subfigure}{\linewidth}
    \centering
    \includegraphics[width=.85\linewidth]{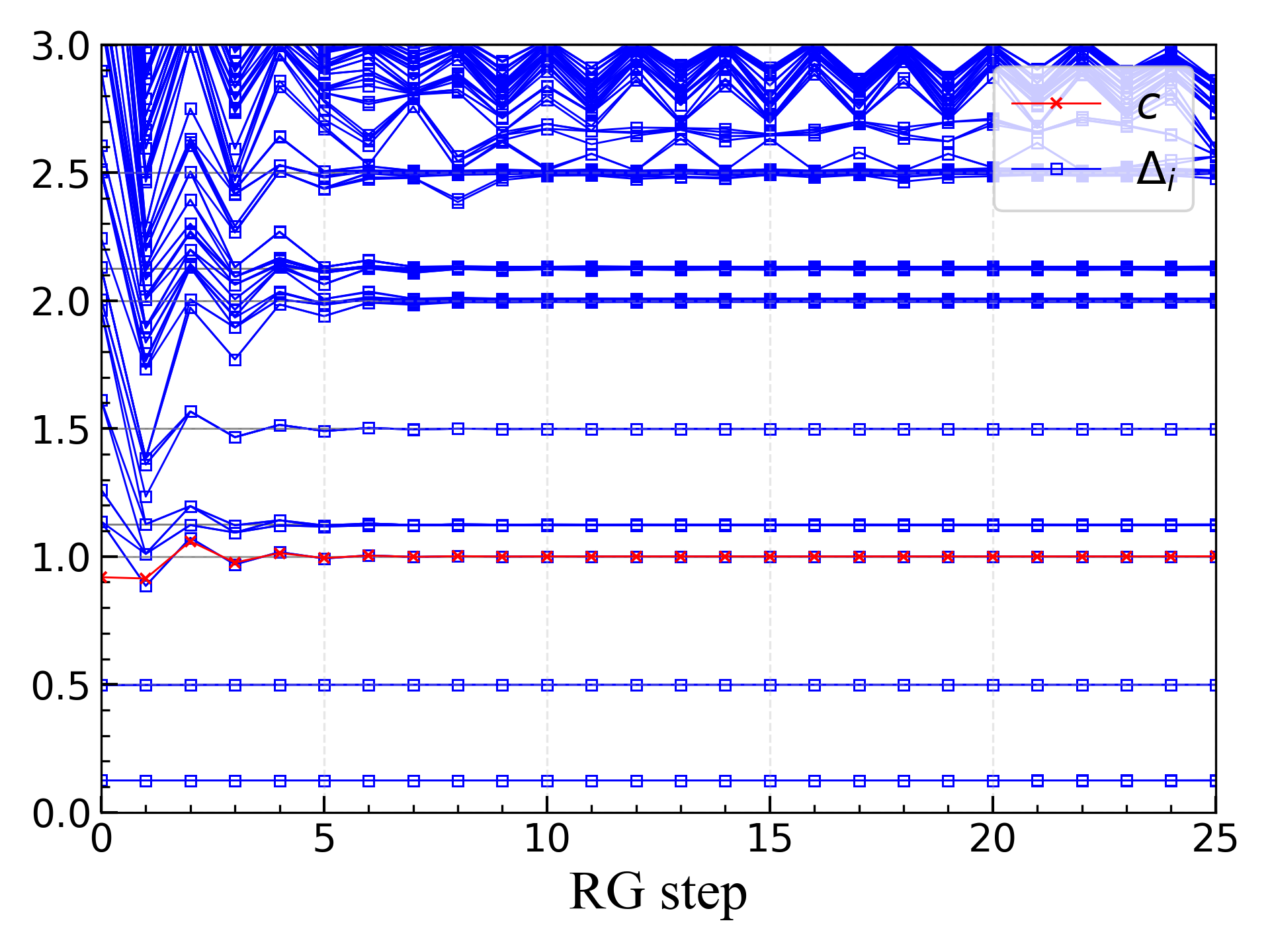}
    \caption{$R=\sqrt{2}$, $\Lambda_D=5$.}
    \label{fig:4pt_D_disk_1.414}
\end{subfigure}
\caption{RG spectrum from the rank-4 FP tensor, $w_{\max}=2$, $\chi=20$.}
\label{fig:4pt_D_disk_123}
\end{figure}

\begin{table}[tb]
\vspace{2em}
\caption{
Central charge and low-lying scaling dimensions for $R=1/\sqrt{2}$ and $\sqrt{2}$, initialized with the rank-4 FP tensor.
}
\label{tab:4pt_D_disk_123}
\centering
\begin{subtable}{\linewidth}
\caption{
$R=1/\sqrt{2}$.
}
\label{tab:4pt_D_disk_0.707}
\centering
\begin{tabular*}{.85\linewidth}{@{\hspace{8pt}} @{\extracolsep{\fill}} c c c @{\hspace{4pt}}}
\hline\hline
\multicolumn{3}{c}{RG steps 10-25: $\quad c_{\rm numerical}=1.0000(2)$} \\
\hline\hline
$\Delta_{\rm exact}$ & degeneracy & $\Delta_{\rm numerical}$ \\
\hline
0.5 & 4  & 0.500(1) \\
1.0 & 6  & 1.0000(2) \\
1.5 & 8 & 1.500(2) \\
2.0 & 17 & 2.001(9) \\
\hline\hline
\end{tabular*}
\end{subtable}
\par\medskip
\begin{subtable}{\linewidth}
\caption{
$R=\sqrt{2}$.
}
\label{tab:4pt_D_disk_1.414}
\centering
\begin{tabular*}{.85\linewidth}{@{\hspace{8pt}} @{\extracolsep{\fill}} c c c @{\hspace{4pt}}}
\hline\hline
\multicolumn{3}{c}{RG steps 10-20: $\quad c_{\rm numerical}=1.0000(2)$} \\
\hline\hline
$\Delta_{\rm exact}$ & degeneracy & $\Delta_{\rm numerical}$ \\
\hline
0.125 & 2  & 0.12461(4) \\
0.5 & 2  & 0.498456(4) \\
1.0 & 2  & 1.0000(3) \\
1.125 & 6  & 1.124(2) \\
1.5 & 4 & 1.4984(3) \\
2.0 & 9 & 2.001(7) \\
2.125 & 18 & 2.126(8) \\
\hline\hline
\end{tabular*}
\end{subtable}
\end{table}

We next investigate the stability of the RG flows generated by the rank-4 FP tensors. The initial rank-4 tensor $\mathcal{T}$ is constructed directly from Eq.~\eqref{eq:forvertex} and Eq.~\eqref{eq:zchi1234} using $\Lambda_D$ Dirichlet CBCs, where again $\Lambda_D$ satisfies the condition in Eq.~\eqref{eq:app_lambdaDN}. The RG spectra for the two radii, (1) $R=1/\sqrt{2}$, and (2) $R=\sqrt{2}$, are shown in Fig.~\ref{fig:4pt_D_disk_123}. Numerical estimates of the central charge and scaling dimensions are summarized in \tabref{tab:4pt_D_disk_123}.

These results demonstrate that the rank-4 FP tensors provide a robust starting point for TCR. Despite being constructed from only a single family of CBCs and in some cases exhibiting slightly larger residual splittings of degenerate levels than the refined rank-3 construction, they substantially suppress the residual marginal perturbation, making them particularly suitable for exploring the exactly marginal line.

\end{document}